\documentclass[12pt]{article}
\usepackage[round,sort,authoryear]{natbib}

\usepackage{amsmath,amssymb,amsthm,amsfonts, fullpage, setspace}

\usepackage{eucal}
\usepackage{subeqnarray,bm}
 \usepackage[utf8]{inputenc}
\usepackage{graphicx}[final]
\usepackage{psfrag} 
\usepackage{color}
\usepackage{pdfpages}

\usepackage{epstopdf}

\usepackage[colorlinks]{hyperref}

\usepackage{hyperref}
\hypersetup{colorlinks=red,citecolor=blue,pdfstartview=FitH, pdfpagemode=None}
\usepackage{multirow}
\usepackage[bottom]{footmisc}
\usepackage{ifthen} 
\usepackage{sectsty}
\usepackage{subfig}
\usepackage{float}
\usepackage[english]{babel}
\usepackage[nottoc]{tocbibind}
\usepackage{calrsfs}
 \DeclareGraphicsExtensions{.eps,.png,.pdf}
\usepackage{mwe}
\makeatletter
\def\maxwidth{ %
  \ifdim\Gin@nat@width>\linewidth
    \linewidth
  \else
    \Gin@nat@width
  \fi
}
\makeatother

\definecolor{fgcolor}{rgb}{0.345, 0.345, 0.345}

\allsectionsfont{\normalsize\bfseries}

\usepackage{framed}
\makeatletter
 {\par\unskip\endMakeFramed%
 \at@end@of@kframe}
\makeatother

\definecolor{shadecolor}{rgb}{.97, .97, .97}
\definecolor{messagecolor}{rgb}{0, 0, 0}
\definecolor{warningcolor}{rgb}{1, 0, 1}
\definecolor{errorcolor}{rgb}{1, 0, 0}

\newtheorem{thm}{Theorem}

\newtheorem{defn}[thm]{Definition}

\newtheorem{remark}[thm]{Remark}

\newcommand{\ds}{\displaystyle}

\newcommand{\norm}[1]{\left\Vert#1\right\Vert}
\newcommand{\abs}[1]{\left\vert#1\right\vert}
\newcommand{\set}[1]{\left\{#1\right\}}

\newcommand{\rb}[1]{\left(#1\right)}

\newcommand{\R}{\mathbb{R}}

\newcommand{\pa}{\partial}

\numberwithin{equation}{section}

\usepackage{calrsfs}

\allsectionsfont{\normalsize\bfseries}

\numberwithin{equation}{section}
\allowdisplaybreaks

\DeclareMathAlphabet{\pazocal}{OMS}{zplm}{m}{n}
\begin{document}
\title{Strong Allee effect synaptic plasticity rule in an unsupervised learning environment}
\author{Eddy Kwessi\footnote{Corresponding author: Department of Mathematics, Trinity University, 1 Trinity Place, San Antonio, TX 78212, Email: ekwessi@trinity.edu}}
\date{}
\maketitle
\begin{abstract}
Synaptic plasticity or the ability of a brain to changes one or more of its functions or structures has generated and is sill generating a lot of interest from the scientific community especially neuroscientists. These interests especially went into high gear after  empirical evidences were  collected  that challenged the established paradigm that human brain structures and functions are set from childhood and only modest changes were expected beyond.  Early synaptic plasticity rules or laws to that regard  include the basic Hebbian rule that proposed a mechanism for  strengthening or weakening of synapses (weights) during learning and memory. This rule however  did not account from the fact that weights must  have bounded growth  overtime. Thereafter, many other rules were proposed to complement the basic Hebbian rule  and  they  also posses   other desirable properties. In particular, a desirable property in synaptic plasticity rule is that the ambient system must account for inhibition which is often achieved  if the rule used allows for a lower bound in synaptic weights.  In this paper, we propose a synaptic plasticity rule inspired from the Allee effect, a phenomenon often observed in population dynamics.  We show  properties such such as synaptic normalization, competition between weights, de-correlation potential,  and dynamic stability are  satisfied. We show  that in fact, an Allee effect in synaptic plasticity can be construed as an absence of plasticity. 
\end{abstract}

\section{Introduction}

Synapses play an important role in the brain  because they are  junctions between nerves cells.  As such, they  facilitate  diffusion of chemical substances called neurotransmitters from the brain to other parts of the body. During this diffusion, synapses are sometimes modified to adapt to the impulses and their transmission rate. These synaptic modifications may be due to lived experience and training and can occur at functional and structural levels. At a functional level, the brain may move functions from one area to other areas, often between damaged to undamaged ones. At a structural level, the brain may actually change its physical structure, mainly some synaptic structures as a result of external activities. Synapses modifications can in turn affect behavior and  training, therefore, understanding the dichotomy between  synaptic modifications and experience and/or  training  is paramount if one wants to have an insight into some of our brain activities. Brain plasticity or neuroplasticity  can be thought of as the ability of the brain to adapt to external activities by reorganizing some of its pathways or by modifying some of its synaptic  structure. Changes in the brain were believed to occur only during infancy and early childhood so that the brain structure was  believed to be mostly set by adulthood. Pioneers challenging this paradigm include \cite{James1890}, who suggested that in fact, at any age, the brain, just like  any organic matter, has a great deal of plasticity.  Early researchers on  synaptic plasticity include  \cite{Hebb}, who conjectured that on one hand,  synapses from two neurons are often strengthened if impulses from one neuron contribute to the firing of another.  On the other hand, synapses are weakened  if non-coincidental neuronal firings occur. In essence, this rule infers that synaptic modifications are in direct relationship with experience and training,  and consequently, the mechanisms underlying learning and memory can be understood via these synaptic modifications. In fact, there is ample empirical evidence consisting of transient and long  lasting effects (long-term potentiation and depression) starting with  \cite{Bliss1973} who experimented plasticity in rabbits.  Plasticity was later experimented in selected regions of the brains like the hippocampus neocortex and cerebellum, see for instance \cite{Bliss1973,Xu2005, Siegelbaum1991, Feldman2009, Bear1994, Liu2018a, Liu2018,Olson2006,Bussey2011}. Learning and memory are therefore manifestations of   the brain's capacity to sustain recurrent  changes. Moreover,  it has been documented that the main  excitatory neurotransmitter in the mammalian nervous system is L-glutamate, see \cite{Voglis2006}. This transmitter is usually detected at postsynaptic terminals by a coupling of G-protein and inotropic glutamate receptors who have been associated with learning and memory, see \cite{Bliss1997}.  Plasticity also plays a role in various pathologies. For instance, following a stroke, evidence has  emerged linking   pathological neural plasticity or postischemic long-term potentiation  with ischemia  which  is a  deficiency of blood to the heart or to the brain,  see \cite{Wang2014}.  Pathological neural plasticity also plays an important role in epilepsy, another pathology that is said to arise from malfunction of a mutated ion channel, leading to increased excitatory or decreased inhibitory currents, see \cite{Wimmer2009, Swann2009}. Other neurological disorders have been linked with plasticity, for instance Alzheimer's disease, Parkinson's disease, Huntington's disease, dystonia, see \cite{Thickbroom2009, Dorszewska2020}. 
However, to understand plasticity at the functional level, one needs to go beyond mechanistic models as described above and find how plasticity relates neurons and/or network of neurons to the basic rules that govern its induction, see  \cite{Dayan2001}. This entails finding mechanisms relating the strengthening or weakening of synapses via neurotransmitters and (presynaptic) neurons. Many mathematical models or synaptic plasticity rules have been proposed to explain synaptic plasticity in supervised and unsupervised learning environments. In an unsupervised learning environment where the neurons network self-organizes, an activity is represented by a continuous variable (input) at the presynaptic level and linked to a postsynaptic activity variable (output) by dynamic weights. The relationship between these variables is a differential equation describing the change of weights overtime and include and is not limited to  the Basic Hebbian rule  (\cite{Sejnowski1989}) and its variant the Covariance rule (\cite{Dayan2001}),  the Bienestock-Cooper-Munro (BCM) rule (\cite{Bienenstock1982}),  and the Oja rule (\cite{Oja1982}). 
To avoid unbounded growth, an upper saturation limit is often imposed, for instance in BCM and Oja rules.  A lower limit is needed to allow for inhibition. However, this lower limit is often given by the condition that the length of weights not be zero. In populations dynamics, there are rules  for which the density or size of a population  is both bounded above and below by nonnegative constants as in the Allee effect. The Allee effect was introduced by  \cite{Allee1949} and characterizes a phenomenon in population dynamics where there is a positive correlation between a  population density or size and its per capita growth rate. In the literature, Allee effects are  divided into strong and weak Allee effects, see for instance \cite{Hutchings2015}. The strong Allee effect occurs when a population has a critical density $A$ below which it declines to extinction while the weak Allee effect occurs when a population lacks such a critical density, but at lower densities, the population growth rate arises with increasing densities. Since their inception, Allee effects have substantially been  investigated and applied by researchers across the board. Ecology is probably the area where researchers have investigated it  the most. Mathematical models of the Allee effects and their dynamics have been investigated for competing populations  in  \cite{Kwessi2014_2, Kwessi2015_8, Kwessi2015_7, Kwessi2015_6, Kwessi2018_4}. Stochastic models of the Allee effects were discussed in \cite{Kwessi2016_3}. Models addressing Allee effects and conservation are  discussed  in \cite{Courchamp2008}.  Models addressing population resilience were proposed by  \cite{Kwessi2015_2}. Some real life evidence of Allee effects have been documented  in \cite{Courchamp2008, Perala2017}. Possible extension of Allee effects to medicine have been proposed in \cite{Delitala2020, Neufeld, Fontanari20061, Konstorum2016, Johnson}. It has been speculated that  the passenger pigeon, who was once the must abundant bird in North America and now extinct,  was subject to  a phenomenon similar  to an Allee effect, see for instance \cite{Fuller2014, Greensberg2014, Avery2014}. The endangered Vancouver Island marmots,  who are on the brink of extinction, may be subject to an Allee effect as well, see for instance \cite{Brashares}.

\begin{figure}[H] 
\begin{flushleft}
   \begin{tabular}{ccc} 
   (a) & (b) & (c)\\
   \includegraphics[width=2.2in]{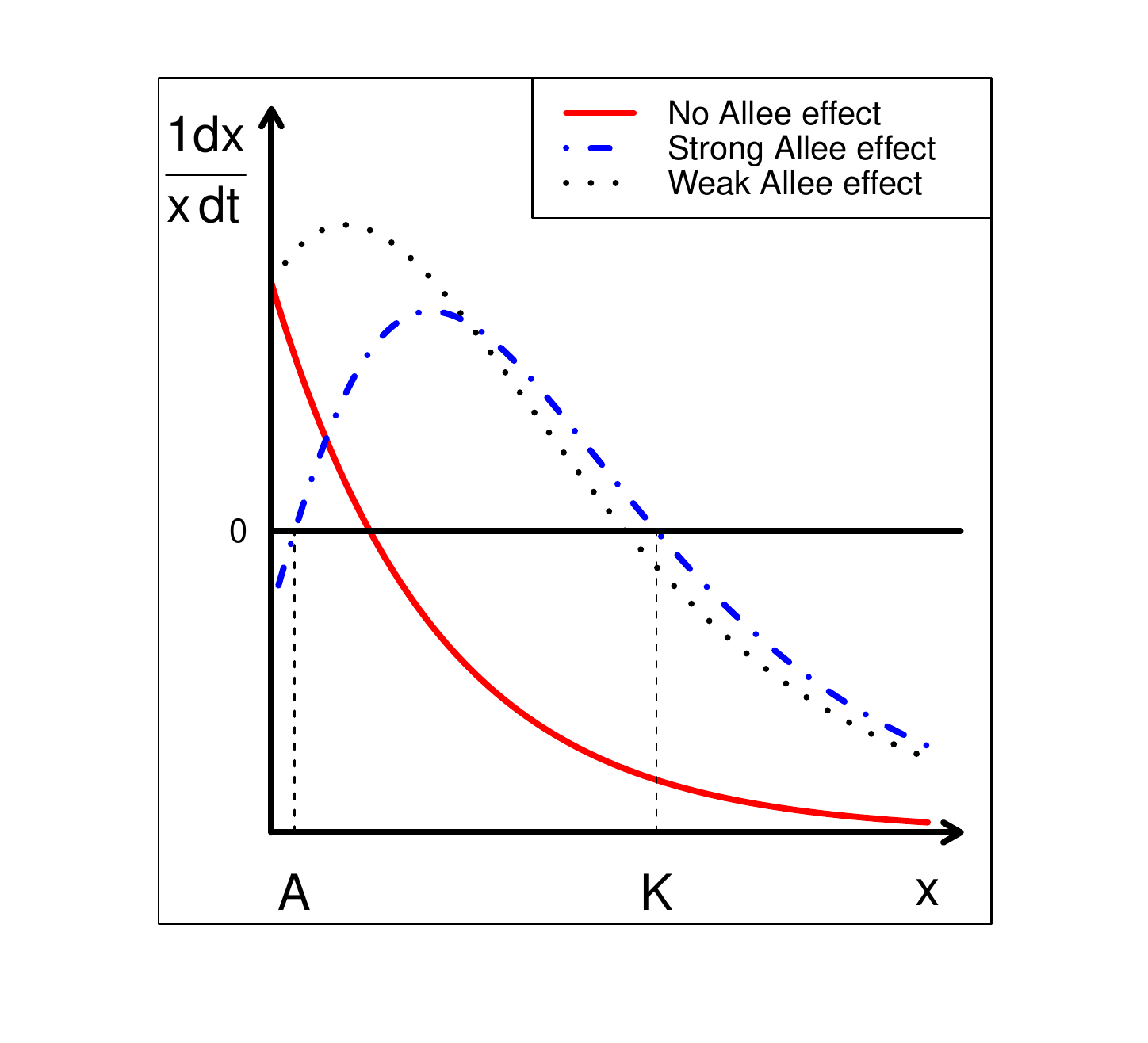}  &  \includegraphics[width=2.2in]{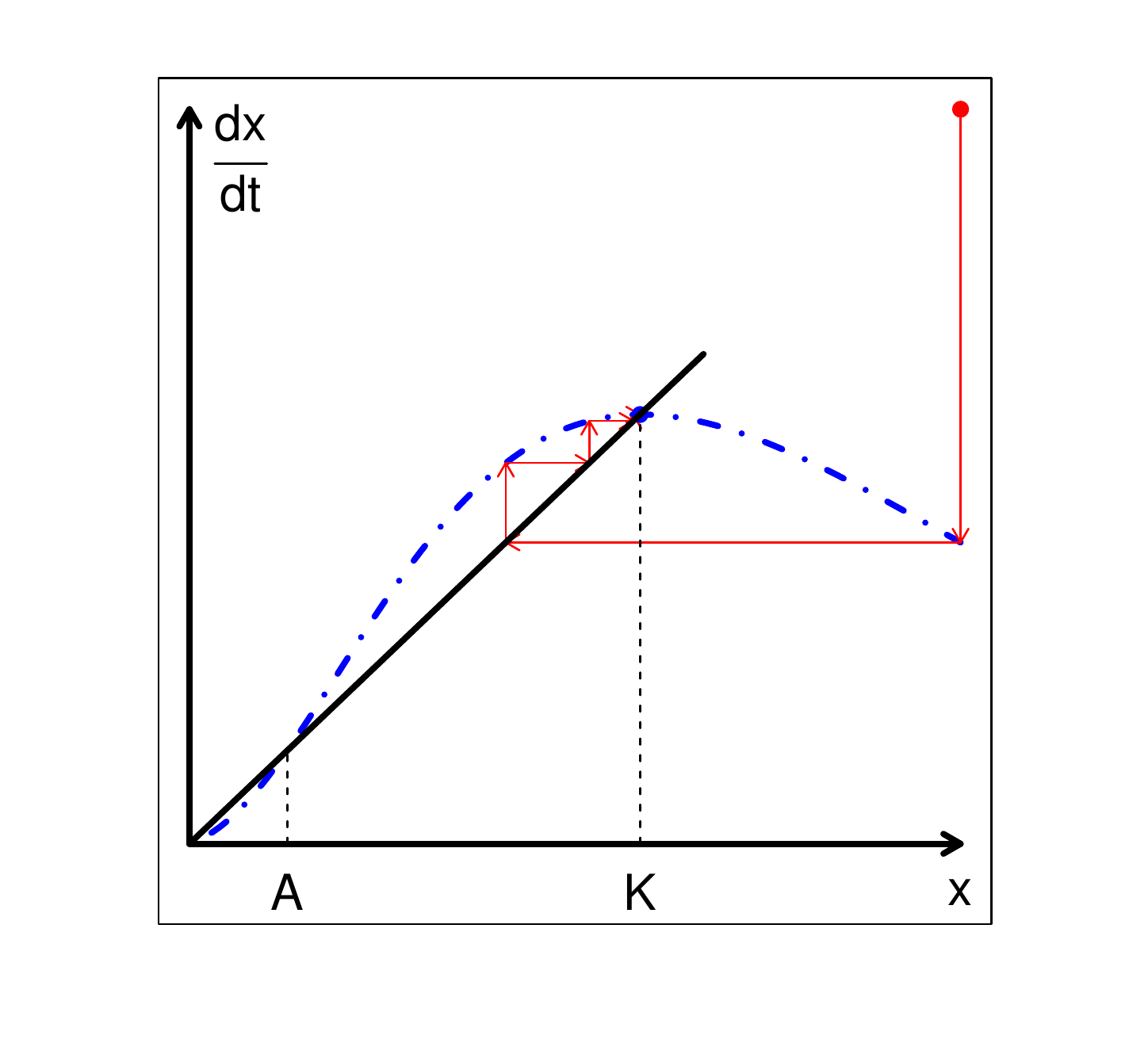} &\includegraphics[width=2.2in]{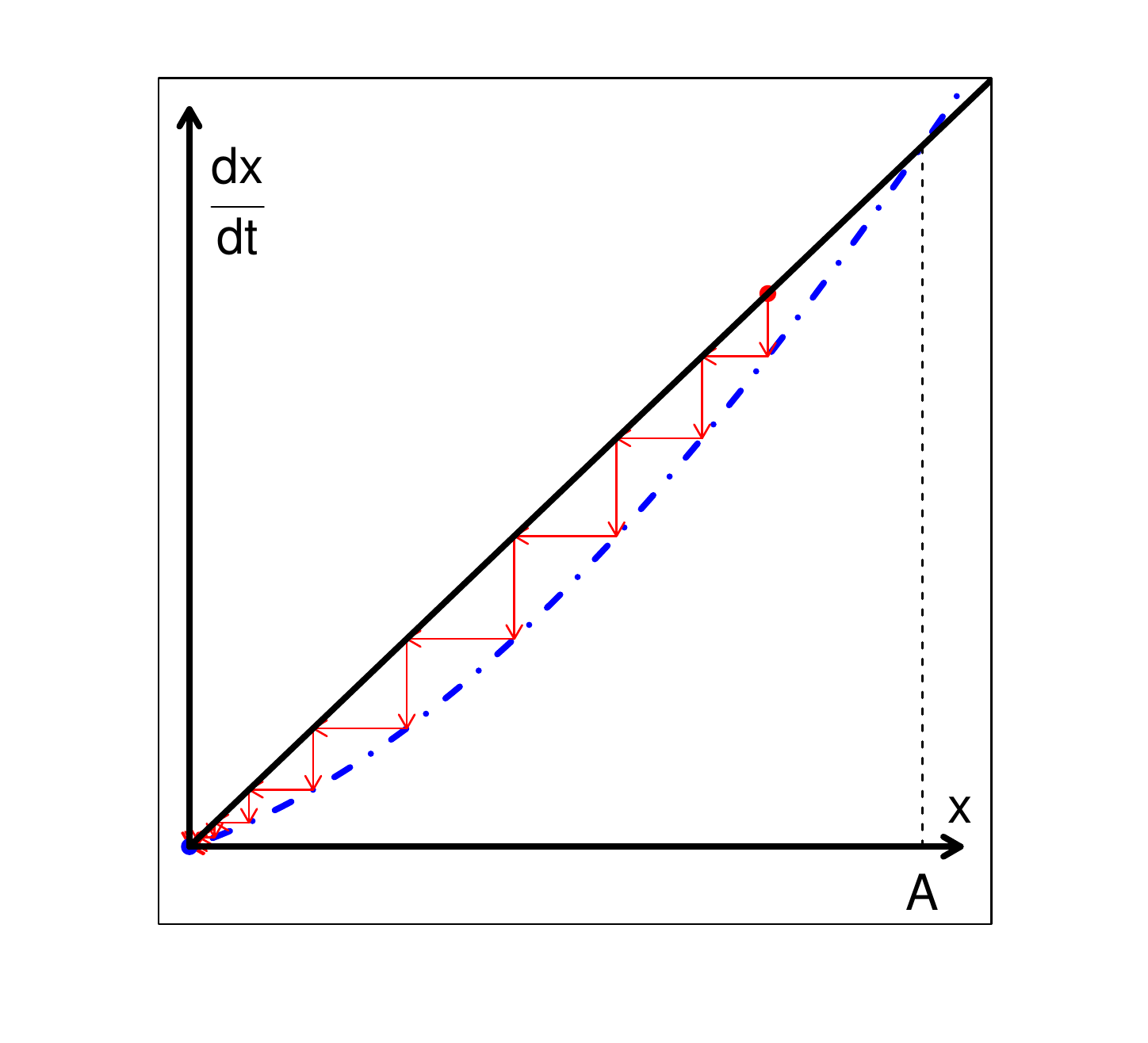} 
   \end{tabular}
   \end{flushleft}
   \caption{ In (a), represented is  the per capita growth rate  as a function of the  population  density  $x$. The red curve represents no Allee effect, the black dashed curve represents the weak Allee effect, blue dashed represents  the strong Allee effect. In (b), the strong Allee effect is represented. A starting trajectory (red arrows) above $A$ converge to $K$. In (c), by zooming   closer $A$, we observe that a starting trajectory below  $A$ converges to 0.}
   \label{fig:AlleeEffect1}
\end{figure}

Figure \ref{fig:AlleeEffect1} below is an illustration of  the different types of Allee effects, where $A$ is the critical (Allee) density and $K$ is the carrying capacity of the population.

To our  knowledge, the Allee effect has not yet  been been discussed in  combination  with plasticity rules. 
 In this paper, we aim to make a foray on the topic and we show that in fact, an Allee effect, when combined to the Oja rule, can be characterized as a drift  toward an absence of plasticity.  Moreover, the model we propose has the following key advantages.
\begin{enumerate}
\item Unbounded growth is controlled.
\item Multiplicative normalization is preserved.
\item Competition between weights is induced.
\item The model is general enough to account for multiple layers of pre-and postsynaptic neurons.
\item Under specific conditions on the network parameters, stability of  dynamical system is obtained.
\end{enumerate}
The  use of an Allee effect in neuroscience may have  the potential to produce invaluable information that could highlight hidden features in plasticity and could potentially  enrich the ever growing  literature on the topic.  The remainder of this paper is organized as follows: In Section \ref{sect3}, we introduce our idea of the an Allee effect postsynaptic neuron model. In Section \ref{sect5}, we discuss stability analysis of s single postsynaptic neuron model, with and without a plastic recurrent connection. Ensembles of postsynaptic neurons are tackled in Section \ref{sect6}.
\section{Allee effect postsynaptic neuron model and motivation} \label{sect3}

Consider a system with $L$  layers and let  a $L\times N_u$ matrix  ${\bf u}=\left({\bf u}^{(1)}, {\bf u}^{(2)},\cdots, {\bf u}^{(L)}\right)$ represent  the presynaptic activities in the system. For $1\leq \ell\leq L, ~{\bf u}^{(\ell)}=\left(u_1^{(\ell)},u_2^{(\ell}),\cdots, u_{N_u}^{(\ell)}\right)$ represents presynaptic activities of $N_u$ inputs or neurons within  the  $\ell$th layer of  the system.  Let a $L\times N_v$ matrix ${\bf v}=\left(\bf{v^{(1)}, v^{(2)},\cdots, v^{(L)}}\right)$ be the postsynaptic activities  generated by the presynaptic activities $\bf u$, where ${\bf v}^{(\ell)}=\left(v_1^{(\ell)},v_2^{(\ell)},\cdots, v_{N_v}^{(\ell)}\right)$ represents the postsynaptic activities  of $N_v$ neurons on  the $\ell$th layer. Let ${\bf W}$ be an  input synaptic block-matrix of weights representing the strengths of the synapses from the presynaptic neurons $\bf u$ to the postsynaptic neurons ${\bf v}$. We note that ${\bf W}$ is   an $L\times N_u\times L\times  N_v$ block-matrix with  entries $\left({\bf W}^{(k,\ell)}\right)$,    for $1\leq k, \ell\leq L$. Each block ${\bf W}^{(k,\ell)}$ is a matrix with entries $w_{ij}^{(k,\ell)}$ where $1\leq j\leq N_v$ and $1\leq i\leq N_u$. To account for inter-connections between postsynaptic neurons, we will consider an   $L\times N_v\times L\times N_v$ recurrence block-matrix $\bf Z$ with entries ${\bf Z}^{(k,\ell)}$,  for $1\leq k, \ell\leq L$,   where each entry ${\bf Z}^{(k,\ell)}$ is a matrix  with entries $z_{mj}^{(k,\ell)}$ for $1\leq j,m\leq N_v$.
 \noindent We let $d_w=L\times N_u\times L\times N_v$ and  $d_z=L\times N_v\times L\times N_v$ and we define the length of a vector $\bf W$ in $\R^{d_w}$ as $\norm{\bf W}^2=\bf W^T\cdot W$, where the dot stands for the dot or inner product in $\R^{d_w}$.
\begin{figure}[H]
\begin{center}
\centering \includegraphics[scale=0.65]{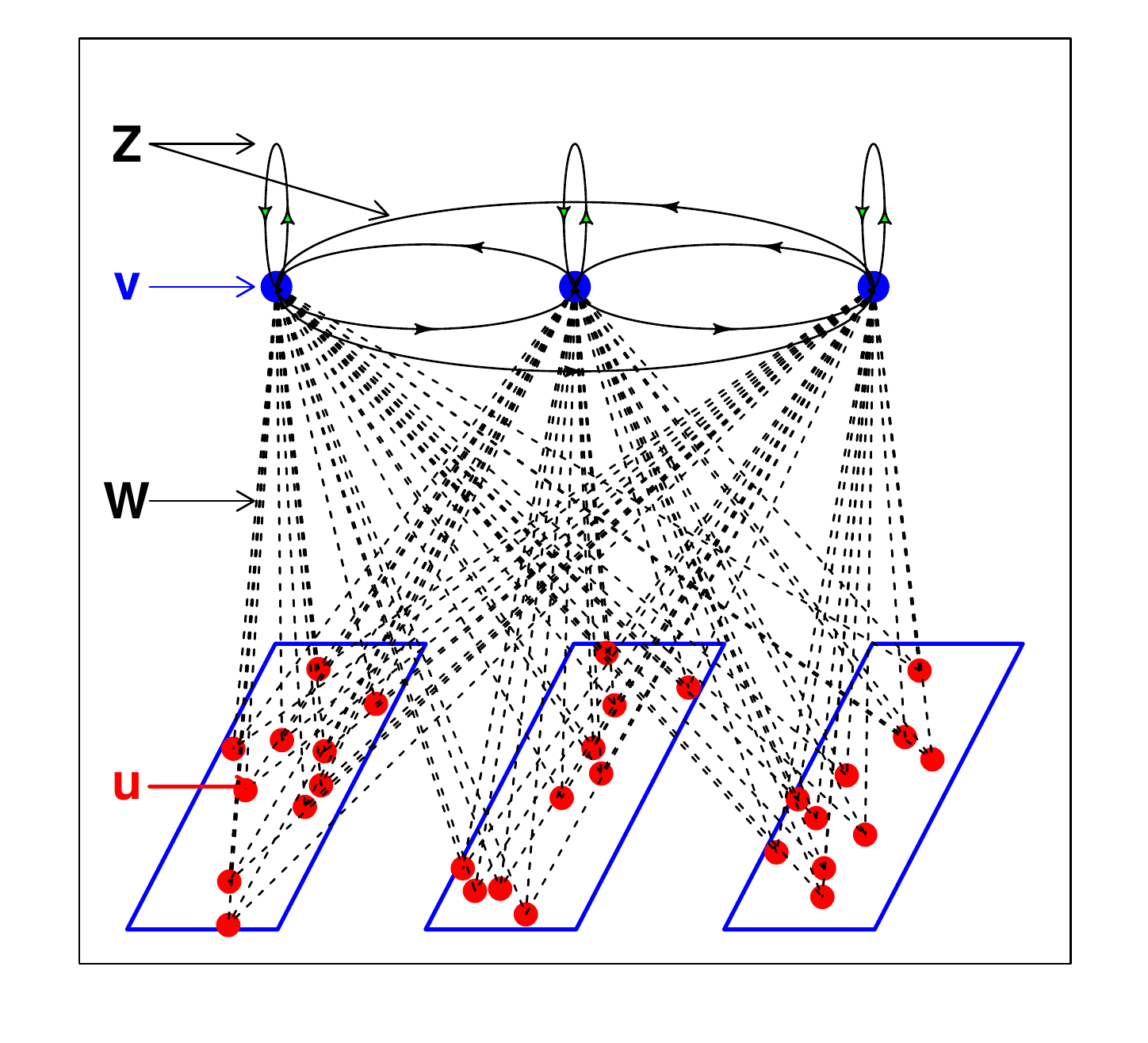}
\caption{A geometric representation  of a triple-layer architecture with $N_u=10$ presynaptic neurons $u_j^{(\ell)}$ and  $N_v=1$ postsynaptic neurons $v_i^{(\ell)}$ per layer, $d_w=90$ weights $w_{ik}^{(k,\ell)}$, and $d_z=9$ recurrent connections $z_{im}^{(k,\ell)}$.}
\label{fig1}
\end{center}
\end{figure}
\begin{remark}\label{rem10}
 In the sequel, we will use invariably the dot product with the following understanding: \\
${\bf W}^T\cdot {\bf u}$ is an $ L\times N_v$ matrix with entries \[({\bf W}^T\cdot {\bf u})^{(k)}_i=\sum_{\ell=1}^L\sum_{j=1}^{N_u}w_{ij}^{(k,\ell)}u_j^{(\ell)},\quad \mbox{for $1\leq k\leq L,~1\leq i\leq N_v$}\;.\]
${\bf Z}^T\cdot {\bf v}$ is an $L\times N_v$ matrix with entries \[({\bf Z}^T\cdot {\bf v})^{(k)}_i=\sum_{\ell=1}^L\sum_{m=1}^{N_v}z_{im}^{(k,\ell)}v_m^{(\ell)}\quad \mbox{for $1\leq k\leq L,~1\leq i\leq N_v$}\;.\]
Consequently, as sums of products of coordinates of vectors, entries for ${\bf W}^T\cdot {\bf u}$ and ${\bf Z}^T\cdot {\bf v}$  are themselves dot products and therefore  enjoy their properties. Also in the sequel,  matrices and vectors and  will be represented by bold face symbols whereas  scalars will be represented by normal font symbols. 
\end{remark}
\begin{defn}
The  learning  activity of a system or  plasticity function of a system  is a function  given as 
\begin{equation}
L({\bf W,u,v})=H({\bf W, u,v})-\varphi({\bf W, u,v})\;,
\end{equation}
where $H({\bf W, u,v})$ is a function referred to as the Hebbian function and $\varphi({\bf W,u,v})$ is  function referred to as the Hebbian modification function.

\end{defn}
\begin{defn} 
We define  a synaptic plasticity rule as 
 
\begin{equation}\label{eq:Hebbrule}
\tau_{_{\bf W}} \frac{d {\bf W}}{dt}=L({\bf W,u,v})\;. 
\end{equation}
\end{defn}
\noindent The constant $\tau_{_{\bf W}}$ represents a time  scaling constant controlling the rate of change of ${\bf W}$ and $\lambda_{\bf W}=\frac{1}{\tau_{_{\bf W}}}$ represents  the learning rate. The model in equation \eqref{eq:Hebbrule} is general enough to include many known synaptic  plasticity rules.  For instance, 
\begin{itemize}
\item If $H({\bf W,u,v})={
\bf v}^T{\bf u}$ and $\varphi({\bf W, u,v})=0$, 
 we obtain the  Basic Hebb rule, see \cite{Hebb}.
 \item If $H({\bf W,u,v})={\bf v}^T{\bf u}$ and $\varphi({\bf W, u,v})=\theta_{\bf v}{\bf u}$ or  $\varphi({\bf W, u,v})=\theta_{\bf u}{\bf v}^T$ for some constants $\theta_{\bf u}$ and $\theta_{\bf v}$, we obtain the Covariance rule, see \cite{Dayan2001}. 
  \item If $H({\bf W,u,v})={\bf v}{\bf v}^T{\bf  u}$ and $\varphi({\bf W, u,v})=\theta_{\bf v}{\bf v}^T{\bf u}$, for some constant $\theta_{\bf v}$, we obtain the Bienestock-Cooper-Munro (BCM) rule, see \cite{Bienenstock1982}.
 \item If  $H({\bf W,u,v})={\bf v}^T{\bf u}$ and $\varphi({\bf W, u,v})=\alpha {\bf v}^T{\bf W}{\bf v}$, for some positive constant $\alpha$, then we obtain the Oja rule, see \cite{Oja1982}.   

 \end{itemize}

 \noindent To model the dynamics of postsynaptic neurons ${\bf v}$, we will use the firing-rate equation given as 
 \begin{equation}\label{eqn:firingPostSyn}
\tau_{_{\bf v}} \frac{d \bf v}{dt}=-{\bf v}+T({\bf W, Z, u,v})\;,
\end{equation}
where $\tau_{_{\bf v}}$ represents the time scale of the firing-rate dynamics of $\bf v$ and  $T({\bf W, Z, u,v})$ is a function representing the  total activity in the system.  This activity  may consist of pre-and postsynaptic activities ${\bf u}$ and ${\bf v}$, with feed-forward and/or  feed-backward  connections with intensities (or weights) ${\bf W}$,  with or without  recurrent connections with intensities  ${\bf Z}$. 
 In the  general literature, $T({\bf W, Z, u, v})$ is  taken as  a linear function of the pre-and postsynaptic activities $\bf u$ and ${\bf v}$. That is, $T({\bf W, Z, u,v})={\bf W^T\cdot u}+{\bf Z^T\cdot v}$. It can also be a nonlinear function depending on an activation function $G$ as $T({\bf W, Z, u,v})=G({\bf W}^T \cdot {\bf u}+{\bf Z} ^T\cdot {\bf v})$ or two activation functions $G_1$ and $G_2$ as $T({\bf W, Z, u,v})={\bf W}^T \cdot G_1( {\bf u})+{\bf Z} ^T\cdot G_2({\bf v})$.  
 The activation function controls the rate of signals emitted by presynaptic neurons $\bf u$ and the recurrence rate  of postsynaptic neurons $\bf v$. It is common to use  either    the  sigmoid function $G(x)=(1+e^{-x})^{-1}$ or the Heaviside function $G(x)=0, x<0, G(x)=1, x>1$. We observe however that more general activation functions $G$ can be considered, see for instance \cite{Kwessi2021_1}. We note that a more complete and perhaps more realistic model for equation \eqref{eqn:firingPostSyn} should contain a diffusion term and a less trivial reaction term than ${\bf v}$. For sake of simplicity and to maintain tractability, we will will not do so in the present discussion.
\begin{remark}\label{rem1}
It is important to note that  here,  the total activity, $T({\bf W, Z, u,v})$ would be zero  if the the pre-and and postsynaptic activities ${\bf  u}$ and ${\bf v}$ are canceling each other. This  is obviously the case if  ${\bf u}=0$ and ${\bf v}=0$. From equation \eqref{eqn:firingPostSyn} above, if $T({\bf W, Z, u,v})=0$, then ${\bf v}(t)={\bf v}_0e^{-\lambda_{\bf v}t}$ and thus approaches 0 overtime. Combining the latter with \eqref{eq:Hebbrule}, it follows, for some constant presynaptic inputs {\bf u} and for some generic constant matrix $\Sigma$ that 
\begin{itemize}
\item   ${\bf W}(t)=\frac{\lambda_{\bf W}}{\lambda_{\bf v}}\left[e^{-\lambda_{\bf v}t}{\bf v}_0^T{\bf u}+\Sigma\right]$ for the Basic Hebb rule. 
\item  ${\bf W}(t)=\frac{\lambda_{\bf W}}{\lambda_{\bf v}}\left[e^{-\lambda_{\bf v}t}{\bf v}_0^T({\bf u}-\theta_{\bf v})+\Sigma\right]$ or ${\bf W}(t)=\frac{\lambda_{\bf W}}{\lambda_{\bf v}}\left[e^{-\lambda_{\bf v}t}{\bf v}_0^T(\theta_{\bf u}-{\bf u})+\Sigma\right]$  for the Covariance rule. 
\item  ${\bf W}(t)=\frac{\lambda_{\bf W}}{\lambda_{\bf v}}\left[\left(\frac{e^{-\lambda_{\bf v}t}}{2}{\bf v}_0^T-\theta_{\bf v}\right)e^{-\lambda_{\bf v}t}{\bf v}_0^T{\bf u}+\Sigma\right]$ for the BCM rule.
\item $\ds {\bf W}(t)=e^{{\bf v}_0^T{\bf v}_0\frac{\alpha }{2\lambda_{\bf v}}e^{-2\lambda_{\bf v}t}}\left[{\bf v}_0^T{\bf u}\int e^{-\lambda_{\bf v}t-{\bf v}_0^T{\bf v}_0\frac{\alpha }{2\lambda_{\bf v}}e^{-2\lambda_{\bf v}t}}dt+\Sigma\right]$ for the Oja rule.
\end{itemize}
We can therefore infer that when $T({\bf W, Z, u,v})=0$,  $\bf v$ approaches 0 overtime whereas $\bf W$ approaches a constant $\Sigma$. Moreover, if $T({\bf W, Z, u,v})={\bf W}^T\cdot {\bf u}+{\bf Z}^T\cdot{\bf  v}=0$, the constant $\Sigma$ must be zero or the presynaptic activities $\bf u$ must be zero.

\end{remark}
\begin{remark}\label{rem2}
We also observe that $T({\bf W, Z, u,v})={\bf W}^T\cdot {\bf u}+{\bf Z}^T\cdot {\bf v}$  can be understood as the total potential energy in the system. Indeed, suppose ${\bf W, Z, u}$, and ${\bf v}$ are 
vector fields over an open connected domain $\mathcal{D}$. Let  $P$ be a path  in $\mathcal{D}$. If these fields are continuous  over  $\mathcal{D}$ and  $\ds \int_P\bf{W}^T\cdot d{\bf u}$ and  $\ds \int_P{\bf u}\cdot d{\bf W}^T$ are path-independent, then  the fields ${\bf W}$ and ${\bf u}$ are conservative. Consequently,  there exist functions $f_1$ and $f_2$ such that ${\bf W}^T=\nabla f_1$ and ${\bf u}=\nabla f_2$. Therefore,  the potential energy due to the fields ${\bf W}^T$ and ${\bf u}$ is   
\[\ds \int_P \nabla f_1\cdot d{\bf u}+\int_P \nabla f_2\cdot d{\bf W}={\bf W}^T\cdot {\bf u}\;.\] 
Similarly,  there exist functions $g_1$ and $g_2$ such that the potential  energy due to the fields ${\bf Z}$ and ${\bf v}$ is \[\ds \int_P \nabla g_1\cdot d{\bf v}+\int_P \nabla g_2\cdot d{\bf Z}^T={\bf Z}^T\cdot {\bf v}\;.\] 
From equation \eqref{eqn:firingPostSyn}, we can deduce that the steady state  is attained when the postsynaptic activity is equal to the total potential energy in the system. This also suggests that postsynaptic activity increases if  it is  less than the system's potential energy and  decreases otherwise. 
\end{remark}

In this paper, we will make the following considerations : for nonnegative constants $A, K$, in equation \eqref{eq:Hebbrule} , we put
\[ 
\begin{aligned}
H({\bf W,u,v})&=\left[{\bf 1}-A ({\bf W}^T{\bf W})^{-1}\right]{\bf v}^T{\bf u}\\
\varphi({\bf W, u,v}) &= K^{-1}\left[{\bf 1}-A ({\bf W}^T{\bf W})^{-1}\right]{\bf v}^T {\bf W}{\bf v},
\end{aligned} 
\]
 where ${\bf 1}$ is a matrix with entries  ones and with the  same dimension as the matrix $({\bf W}^T{\bf W})^{-1}$. Equation \eqref{eq:Hebbrule} can  now be written as 
\begin{equation}\label{eq:Allee1}
\tau_{_{\bf W}} \frac{d {\bf W}}{dt}={\bf v}^T\left({\bf u}-K^{-1}{\bf  W}{\bf v}\right)\left({\bf 1}-A ({\bf W}^T{\bf W})^{-1}\right)\;.
\end{equation}
Let us now give a motivation for the model in equation  \eqref{eq:Allee1}.
 Suppose that we have only one layer($L=1$) and a single postsynaptic neuron ($N_{v}=1$). Therefore ${\bf v}=v$ will be a  scalar.  Let us assume further that there is  no recurrent connection ($\bf Z=0$). Per Remark \ref{rem10}, ${\bf W}^T\cdot {\bf u}$ will be  a scalar. Moreover, ${\bf W}$ is a $1\times N_u$ vector, thus ${\bf W}^T\cdot {\bf W}=\norm{\bf W}^2$ is a positive scalar. Equation \eqref{eq:Allee1} becomes
 \begin{equation}\label{eqn:Allee01}
 \tau_{_{\bf W}} \frac{d {\bf W}}{dt}=v\left({\bf u}-K^{-1}v{\bf  W}\right)\left({\bf 1}-A (\norm{\bf W}^2)^{-1}\right)\;.
 \end{equation}
 Thus, if we take the dot product on both sides of equation \eqref{eq:Allee1} by ${\bf W}^T$ and multiply by the constant 2,  it becomes  
\begin{equation}\label{eqn:Alleweightequare}
2\tau_{_{\bf W}}{\bf W}^T\cdot \frac{d{\bf W}}{dt}=\tau_{_{\bf W}}\frac{d\norm{\bf W}^2}{dt}=2v\left({\bf W}^T\cdot {\bf u}-\frac{v}{K}\norm{\bf W}^2\right)\left(1-\frac{A}{\norm{\bf W}^2}\right)\;.
\end{equation}
\noindent Consequently, the system composed of equations \eqref{eqn:Alleweightequare} and  \eqref{eqn:firingPostSyn}  arrives at its steady states, given by the equation of its $v$-isocline $\left( \ds \frac{dv}{dt}=0\right)$  $v=T({\bf W,u},v)$   and  the equations of its $\norm{\bf W}^2$-isocline $\left(\frac{d\norm{\bf W}^2}{dt}=0\right)$ given as $v=0, \norm{\bf W}^2=A$, and ${\bf W}^T\cdot {\bf u}-\frac{v}{K}\norm{\bf W}^2=0$. In the linear case in particular, we have   $ v=T({\bf W,  u},v)={\bf W}^T\cdot {\bf u}$. Therefore, we obtain that either $\norm{\bf W}^2=A$ or $\norm{\bf W}^2=K$ when the steady states are reached. Similarly to  a strong Allee effect,  the lengths of weights  $\norm{\bf W}^2$ below $\min\set{A,K}$ will   decay to zero overtime, which means that  postsynaptic activities  will decrease towards a state of no activity at all.  We note also that when $A=0$, equation \eqref{eq:Allee1} becomes the Oja rule. In population dynamics  in general, the term $\left(1-A\norm{\bf W}^{-2}\right)$ is used to account for the presence of sparse populations or mate limitation. Adding this term as in equation \eqref{eqn:Alleweightequare} induces the synaptic normalization in that when the weights are nonnegative, their growth is limited by the global treshholds $A$ or $K$, see Figure \ref{fig:PSD1}--\ref{fig:PSD5} below, in the case of a single neuron. Moreover, convergence of $\norm{\bf W}^2$ toward $K$ induces competition between weights and preserves dynamic stability. In conclusion, the model in equation \eqref{eq:Allee1} is a matrix equivalent of equation \eqref{eqn:Allee01} when we account for multiple layers and multiple postsynaptic neurons. 
In light of these facts, we will combine equation  \eqref{eq:Allee1} and \eqref{eqn:firingPostSyn} for the definition of an Allee plasticity rule.
\begin{defn}
An Allee effect is  a synaptic plasticity rule having   a treshhold under which, activities drift towards an absence of plasticity and above which activities eventually become stable.
\end{defn}
\begin{defn} Let ${\bf W, Z,  u}$, and  ${\bf v}$ be given as above. Consider  an activity function $T({\bf W, Z, u, v})$. We define an Allee plasticity rule with non plastic recurrent connections as the system of differential equations
\begin{equation}\label{eqn:Alleerule1}
\begin{cases}
\ds \tau_{_{\bf W}} \frac{d {\bf W}}{dt} \vspace{0.25cm}&={\bf v}^T\left({\bf u}-K^{-1}{\bf  W}{\bf v}\right)\left({\bf 1}-A ({\bf W}^T{\bf W})^{-1}\right)\\
\ds \tau_{_{\bf v}} \frac{d \bf v}{dt}&=-{\bf v}+T({\bf W, Z, u,v})
\end{cases}\;.
\end{equation}
\end{defn}
\begin{defn}  We define an Allee plasticity rule with  plastic recurrent connections  as the system of differential equations
\begin{equation}\label{eqn:Alleerule12}
\begin{cases}
\ds \tau_{_{\bf W}} \frac{d {\bf W}}{dt}  \vspace{0.25cm}&={\bf v}^T\left({\bf u}-K^{-1}{\bf  W}{\bf v}\right)\left({\bf 1}-A ({\bf W}^T{\bf W})^{-1}\right)\\
\ds \tau_{_{\bf v}} \frac{d \bf v}{dt}  \vspace{0.25cm}&=-{\bf v}+T({\bf W, Z, u,v})\\ 
\ds \tau_{_{\bf Z}} \frac{d \bf Z}{dt}&=R({\bf W, Z, u,v})
\end{cases}\;.
\end{equation}
\end{defn}
\noindent where $R({\bf W, Z, u,v})$ is the total {\it  recurrent (post-synaptic) activity} and $\tau_{_{\bf Z}}$ is a scaling constant. 
In the literature, two rules are often considered for $R({\bf W, Z, u,v})$,  see for instance  see \cite{Dayan2001}. 
\begin{itemize}
\item Anti-Hebbian rule: $R({\bf W, Z, u,v})=-{\bf v}^T{\bf v}+\beta {\bf Z}$, for some constant $\beta$.
\item Goodall rule: $R({\bf W, Z, u,v})={\bf I}-(\bf {\bf W}^T\cdot {\bf u}){\bf v}-{\bf Z}$. This rule is often used because  it produces de-corralated postsynaptic outputs and possess  homoschedasticity properties. 
\end{itemize}
In the next sections, we will discuss stability analysis of these plasticity rules. 
\section{ Stability analysis of a  single postsynaptic neuron model} \label{sect5}
Here, we let $L=1$ and $N_v=1$. In the presence of a single postsynaptic neuron, the matrix ${\bf v}$ is reduced to a constant  $v$ and the recurrent connection matrix $\bf Z$ is reduced to a single constant  $z$. There are two cases to consider: firstly, there could be  no recurrent connection between $v$ and itself ($z=0$), see Figure \ref{fig2} {\bf (a)}  below. Secondly, there could be  a recurrent connection with weight $z\neq 0$, see Figure \ref{fig2} {\bf (b)}  below. 

\begin{figure}[H] 
   \centering
   \begin{tabular}{cc}
   {\bf (a)} & {\bf (b)}\\
    \includegraphics[scale=0.5]{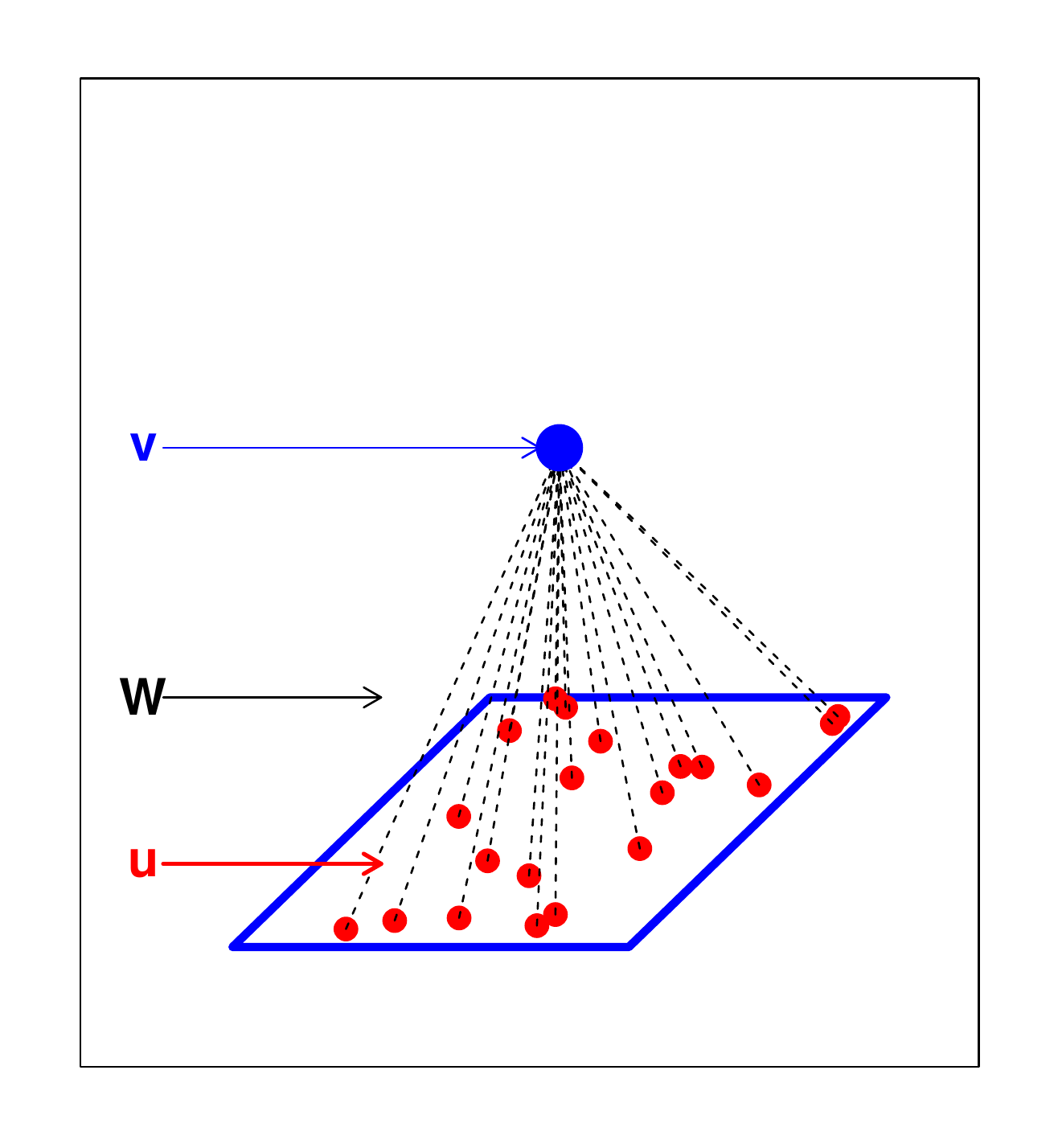} &
   \includegraphics[scale=0.5]{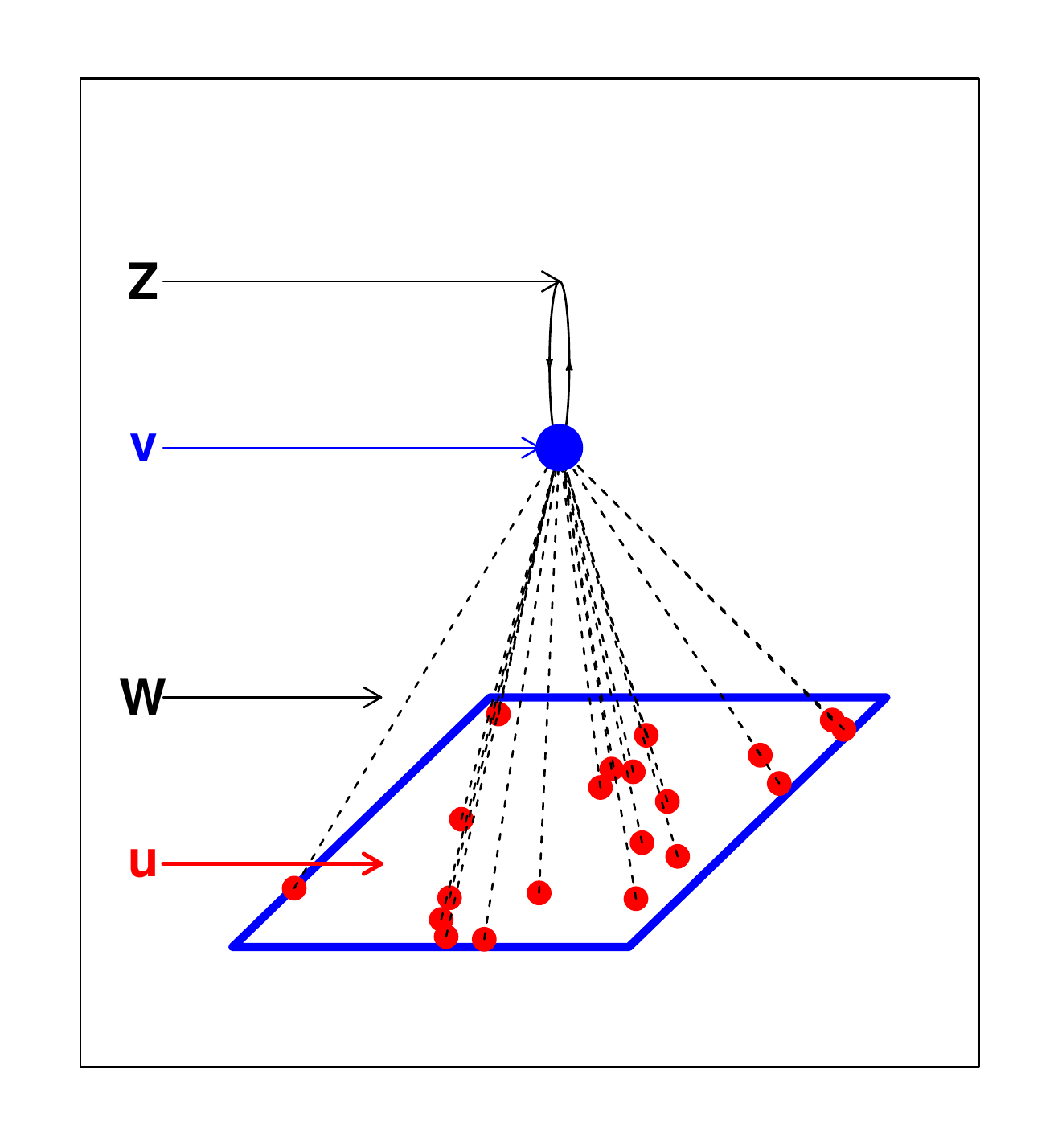}   
   \end{tabular}
      \caption{In (a), we have one postsynaptic neuron with no recurrent connection. In (b), we have one postsynaptic neuron with one recurrent connection.}
   \label{fig2}
\end{figure}

\subsection{Single postsynaptic neuron with no recurrent connection}

In this case, we will have ${\bf Z}=0$ and ${\bf v}=v$. For sake of simplification, we let $T({\bf W}, 0,{\bf u},v)=T(\bf W, u)$. As we observe  above, we consider the scalar auxiliary Allee-type system given by 
\begin{equation}\label{eqn:Alleerule2}
\begin{cases}
\ds \tau_{_{\bf W}}\frac{d\norm{\bf W}^2}{dt}&=\ds 2v\left({\bf W}^T\cdot {\bf u}-\frac{v}{K}\norm{\bf W}^2\right)\left(1-\frac{A}{\norm{\bf W}^2}\right)\;\\
\ds \ds \tau_{v} \frac{dv}{dt}&=-v+T({\bf W, u})
\end{cases}\;.
\end{equation}
\begin{remark}The steady states of this system will be given by the $v$-isocline $v=T({\bf W,u})$ and the $\norm{\bf W}^2$-isocline $v=0, \norm{\bf W}^2=A$, or 
$\ds \frac{{\bf W}^T\cdot {\bf u}}{\norm{\bf W}^2}=\frac{1}{K}T({\bf W, \bf u})$.
The latter equation has a geometric interpretation.  Indeed, this means that  within the sphere with radius $r=\norm{\bf W}$, $\ds  \frac{1}{K}T({\bf W, \bf u})$ is the  {\it scalar projection} of ${\bf u}$ onto the vector $\bf W $ whereas $ \frac{1}{K}T({\bf W, \bf u}){\bf W}$ is  the {\it vector projection} of  ${\bf u}$ onto the vector ${\bf  W}$, see Figure \ref{fig3} below.

\begin{figure}[h!]
\begin{center}
\centering \includegraphics[scale=.6]{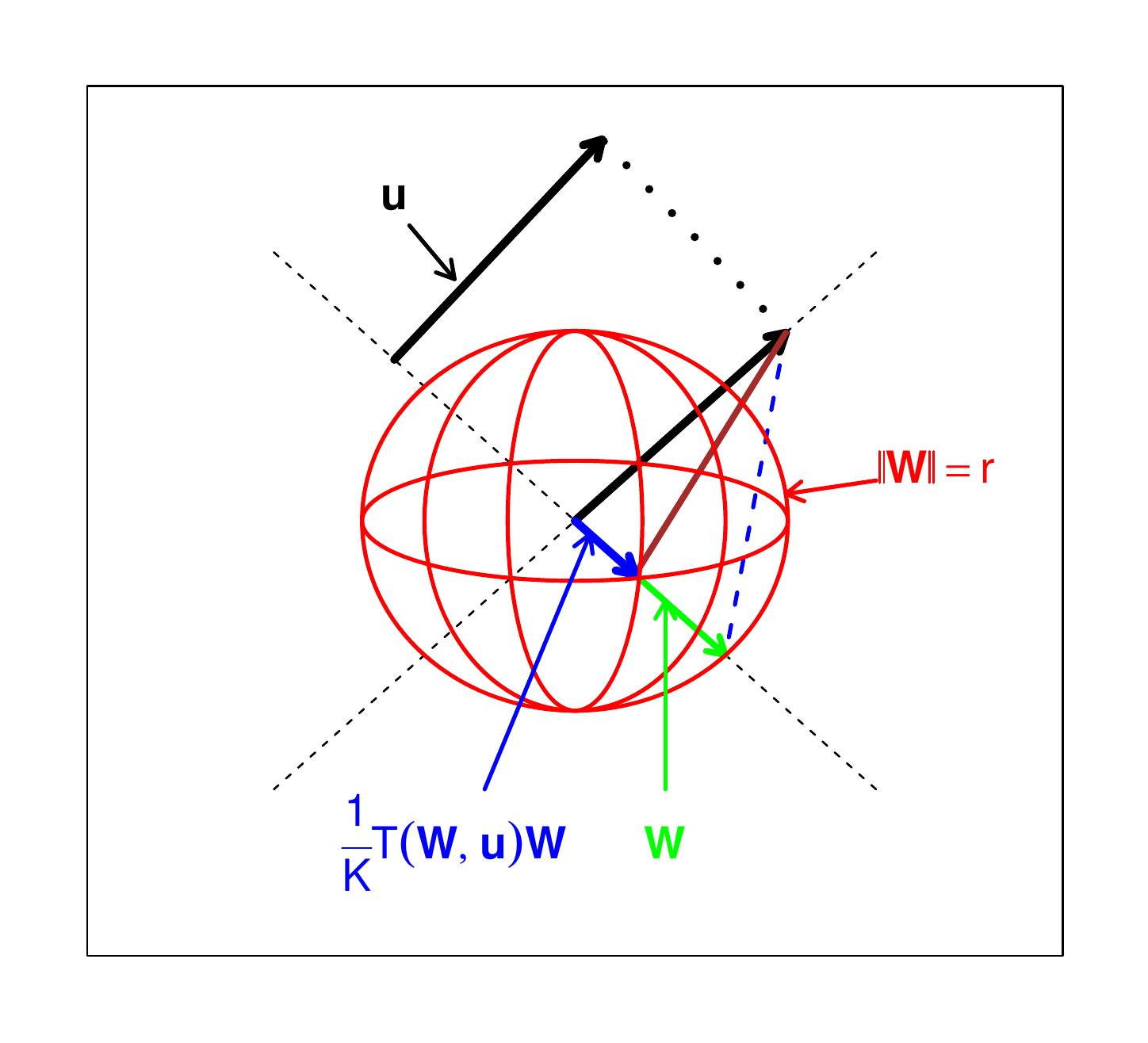}
\caption{Geometric interpretation of $\ds \frac{1}{K}T({\bf W, \bf u}){\bf W}$ when the steady state is attained.}
\label{fig3}
\end{center}
\end{figure}
\end{remark}
We will use the notation $x=:\norm{\bf W}^2, y:=v, u=\norm{\bf u}\cos(\theta) $, where $\theta$ is the angle between the vector $\bf W$ and $\bf u$.   Thus for suitable  functions $f_1(x,u)$ and $f_2(x,u)$,  this system is of the form
\begin{equation}\label{eqn:Alleerule22}
 \begin{cases}
\ds \tau_{x}\frac{dx}{dt} \vspace{0.2cm}&=g_1(x,y):=2y\left(f_1(x,u)-\frac{yx}{K}\right)\left(1-\frac{A}{x}\right)\;\\ 
\ds \tau_y \frac{dy}{dt}&=g_2(x,y):=-y+f_2(x,u)
\end{cases}\;.
\end{equation}
\noindent Let us discuss the local stability of  the steady state $(A,f_2(A,u))$.

\begin{thm}\label{thm1}
Consider the  system  \eqref{eqn:Alleerule22}. Put $a_0=f_1(A,u)$ and $b_0=f_2(A,y,u), \ds c_0:=\frac{\pa f_2(A,u)}{\pa x}$.

\begin{itemize}
\item[(i)] The line $y=0$ for all $x>0$ is always a spiral sink.

\item[(ii)] If $b_0\left(a_0K-b_0A\right)<0$,  then the steady state  $(A,b_0)$  is asymptotically stable.

\item[(iii)] If  $b_0\left(a_0K-b_0A\right)>0$, then $(A,b_0)$  is unstable (repeller).

\item [(iv)]If  $\left(a_0K- b_0A\right)=0$, the steady state $(A,b_0)$  is a saddle.
\end{itemize}

\end{thm}

 \begin{remark}
  \begin{itemize}
  \item[(a)] We note  that the classification of a  steady state is independent of the sign of  $u$ because $\Delta, tr(J)$, and $det(J)$ all depend on $u^2$. 
  \item[(b)] We also note that dynamics of the system  in  the nonlinear case where $T_2({\bf W,u},v)={\bf W}\cdot G(\bf u)$ for some nonlinear function $G$ are  similar to that of the linear case. Special care needs to be taken in the case of the steady state $(x,0)$. This would require, per remark \ref{rem1} above, that  $G(\bf u)=0$. Since $G(x)>0$ for all $x\in \mathbb{R}$, $(x,0)$ cannot be a steady state.  In the case of the  Heaviside activation function, $(x,0)$ is a steady state only when the presynaptic activities are  negative (in inhibition), otherwise, it is not a steady state. 
  \end{itemize}
  \end{remark}
  \noindent In Figure \ref{fig:PSD1} --\ref{fig:PSD5}  below, we will illustrate the above results by plotting the time series  of $x_t$ and $y_t$, for $t=0,\cdots, 250$.  We let $A$ and $K$ take interchangeably the values 1.5 and 3 in Figure \ref{fig:PSD1} , \ref{fig:PSD2} , \ref{fig:PSD4} . The starting points of the trajectories are $(0.1,-2), (0.2,-0.9)$ in  black color, $(0.3,1.1), (0.4,1.5)$ in cyan color, $(4.8,1), (5,2)$ in brown color, and $(9.8,-2), (4.5,-2)$ in magenta color.  In  Figure \ref{fig:PSD1} , \ref{fig:PSD2} , \ref{fig:PSD3} , \ref{fig:PSD5} , we take $u=0.3$. The solid  blue curve represents the $x$-isocline $\ds f_1(x,u)-\frac{yx}{K}=0$ which, in the linear case, is given as $\ds y=\frac{uK}{\sqrt{x}}$. The solid  blue line represents the  $x$-isocline $y=0$ and the solid  red curve represents the $y$-isocline $y=f_2(x,u)$ which, in the linear case,  is given as $y=u\sqrt{x}$. The dots are the intersection between the $x$- and $y$-isoclines. In particular, the gray dot represents  the origin $(0,0)$, the yellow dot has coordinates coordinates $(A, u\sqrt{A})$, and  is the intersection between the $x$-isocline $x=A$ and the $y$-isocline $y=u\sqrt{x}$. Likewise, the green dot is the intersection between the $x$-isocline $x=K$ and the $y$-isocline  $y=u\sqrt{x}$ with coordinates $(K, u\sqrt{K})$. From these figures, we confirm the results above in that $(A, u\sqrt{A})$ and $(K, u\sqrt{K})$ are either attractors or saddle points. Figure \ref{fig:PSD4}  confirms that when $u=0$,  the line $y=0$ is a sink. 
  \begin{figure}[H] 
  \begin{tabular}{cc}
  \includegraphics[scale=.55]{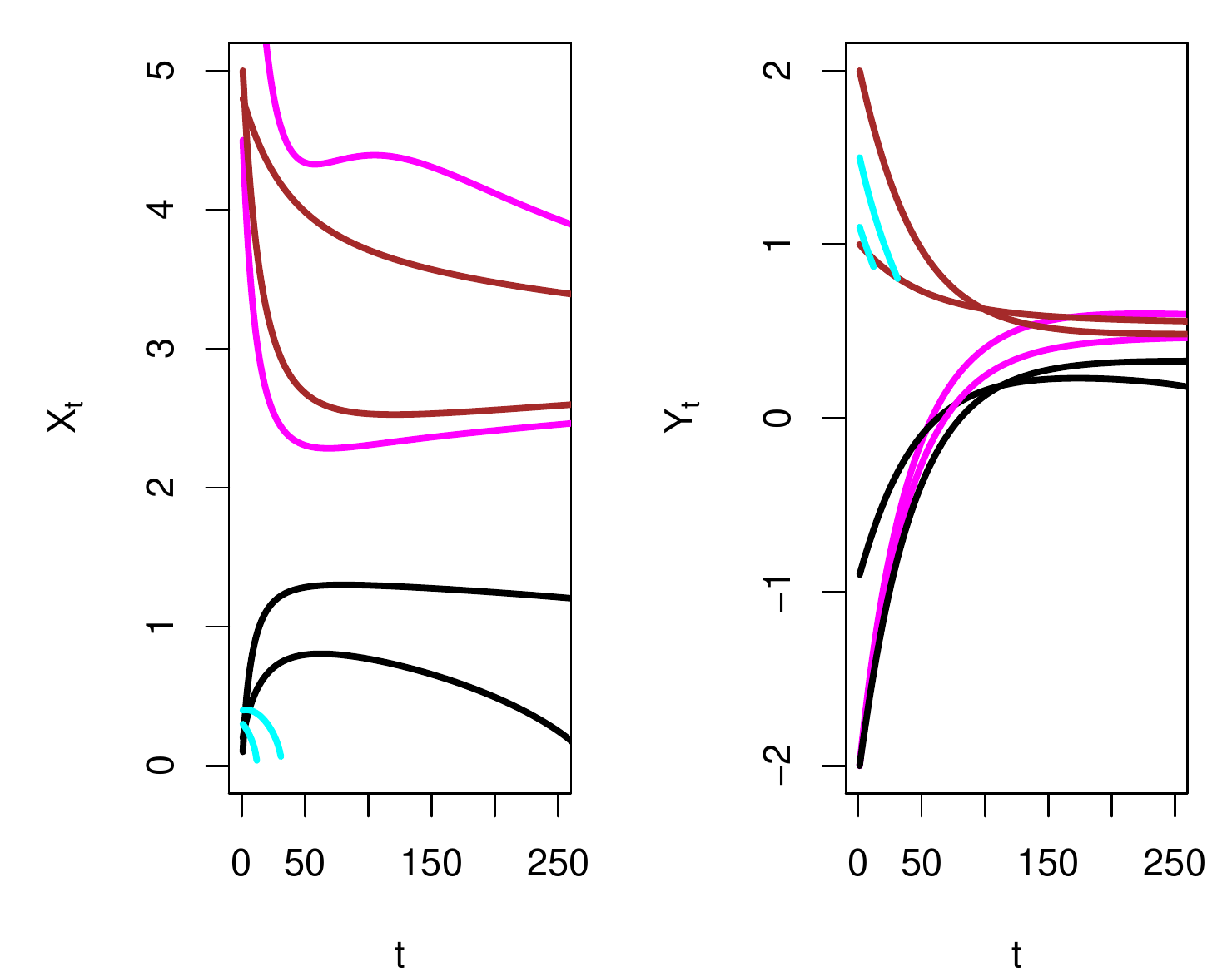}  &  \includegraphics[scale=.55]{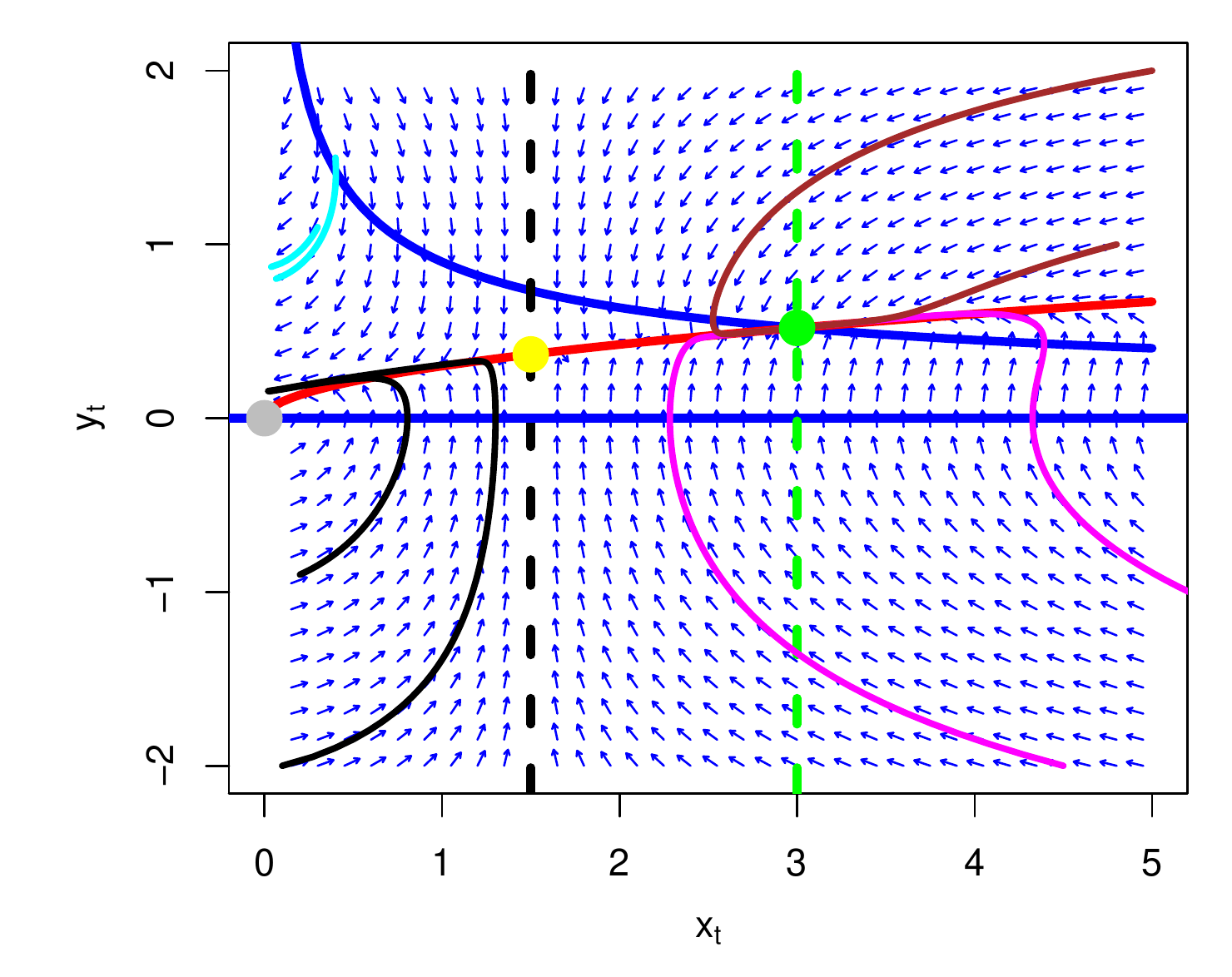}  
    \end{tabular}
    \caption{Time series and phase space diagram when $A=1.5$ and $K=3$. The point $(A, u\sqrt{A})$ is repeller and $(K, u\sqrt{K})$ is an attractor.}
    \label{fig:PSD1}
  \end{figure}
    \begin{figure}[H] 

  \begin{tabular}{cc}
    \includegraphics[scale=.55]{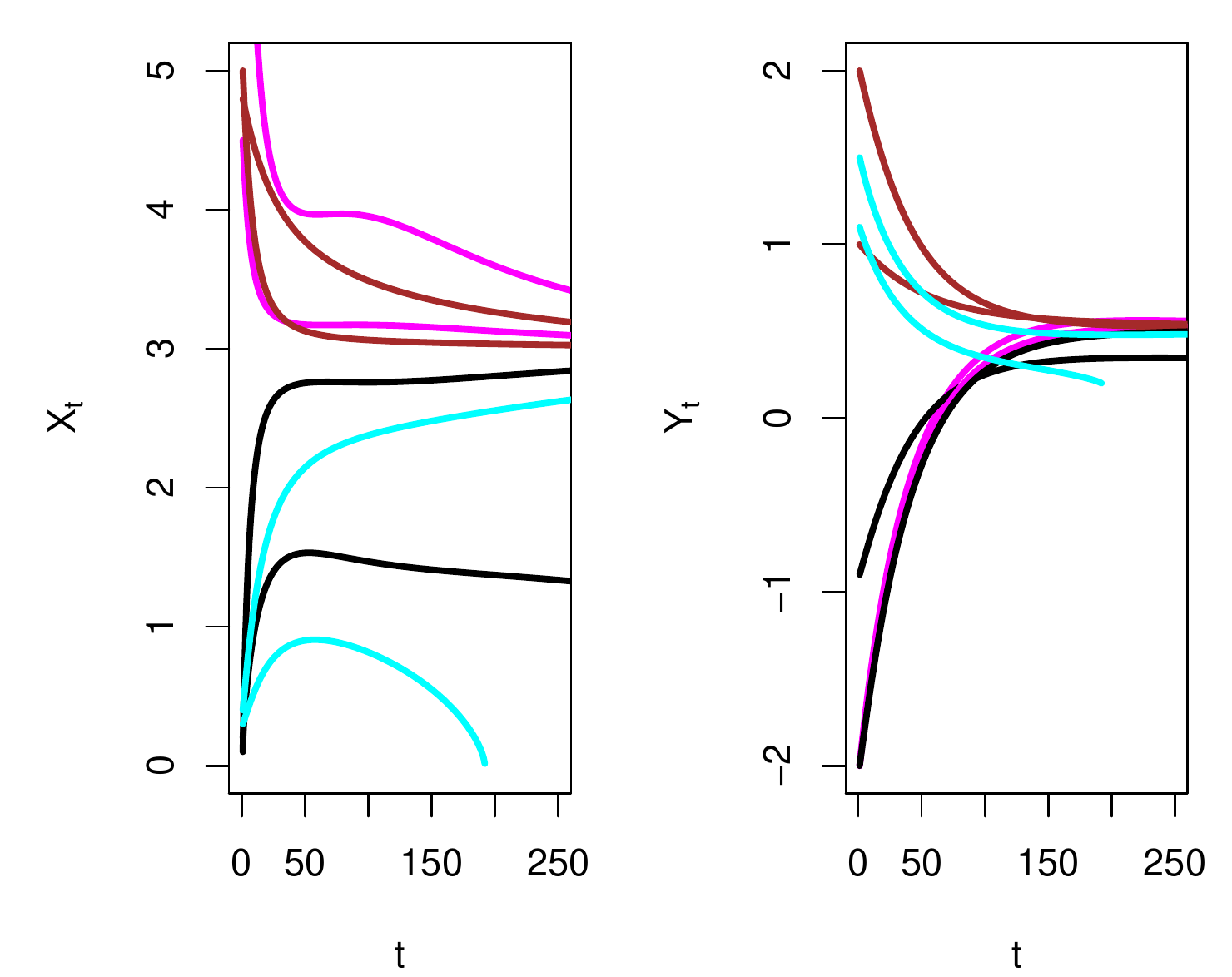}  &  \includegraphics[scale=.55]{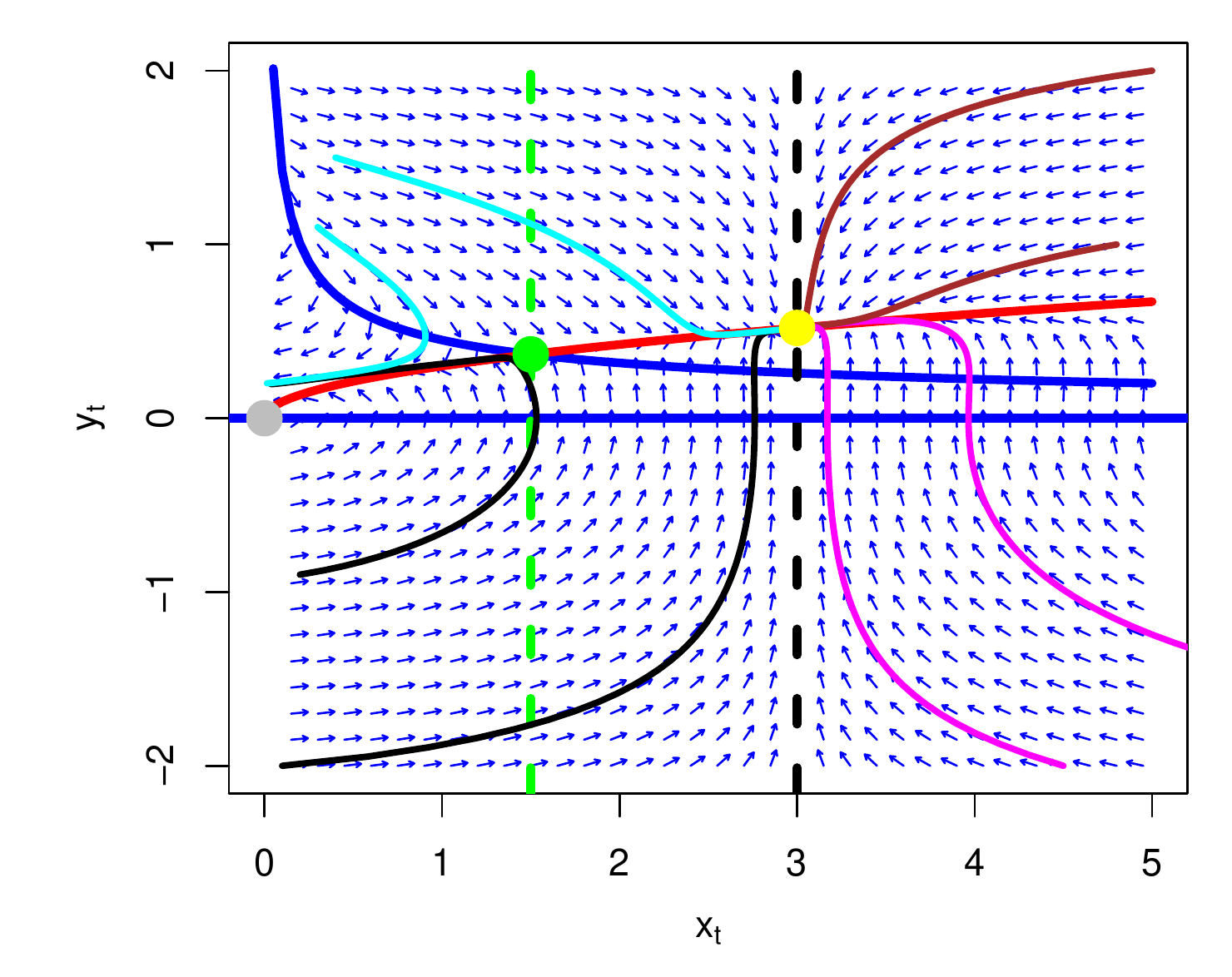}  
      \end{tabular}
       \caption{Time series and phase space diagram when $A=3$ and $K=1.5$. The point $(A, u\sqrt{A})$ is an asymptotically stable (attractor) and $(K, u\sqrt{K})$ is neither an attractor nor a repeller.}
  \label{fig:PSD2}
  \end{figure}
      \begin{figure}[H]  
  \centering
  \begin{tabular}{cc}
       \includegraphics[scale=.55]{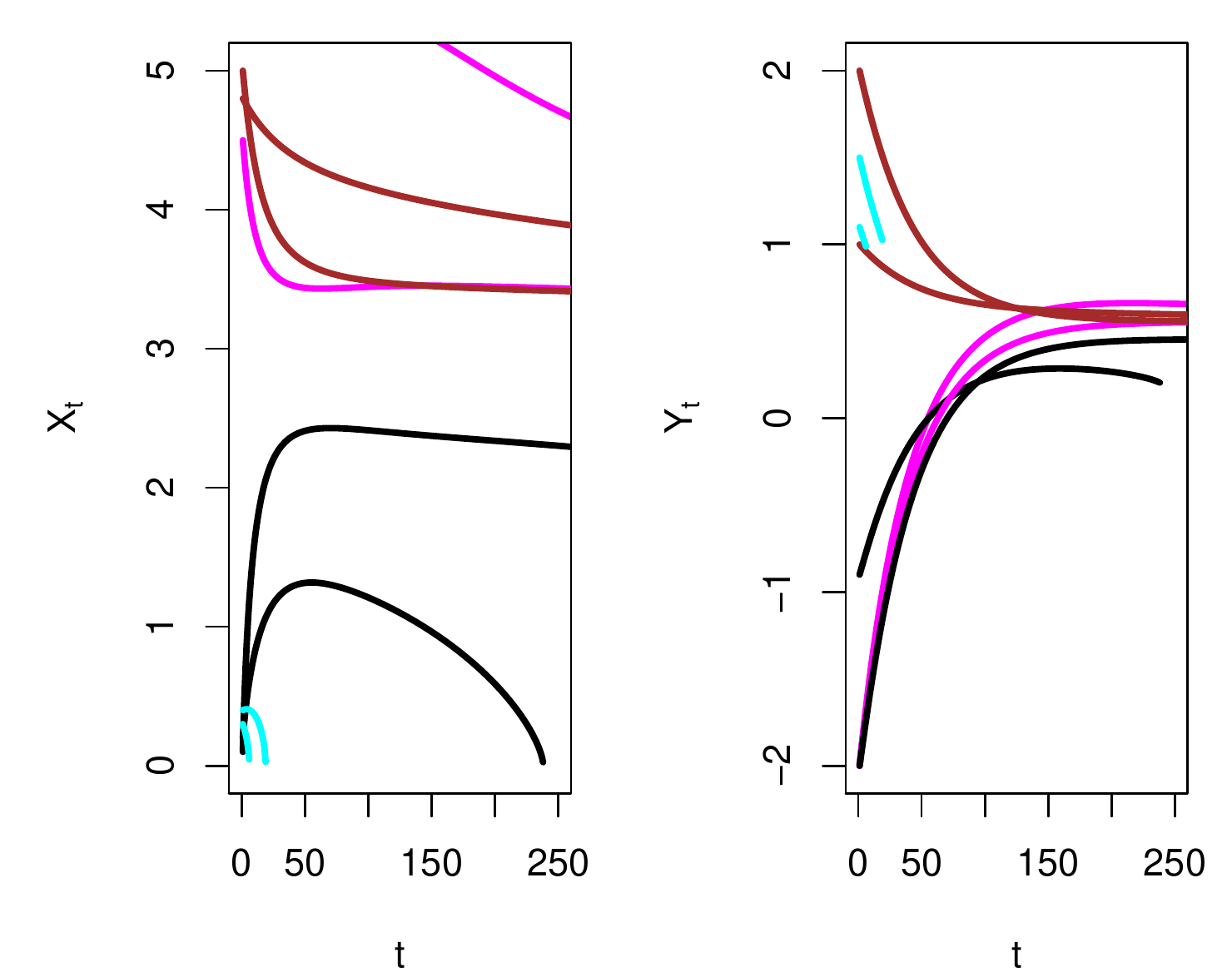}  &  \includegraphics[scale=.55]{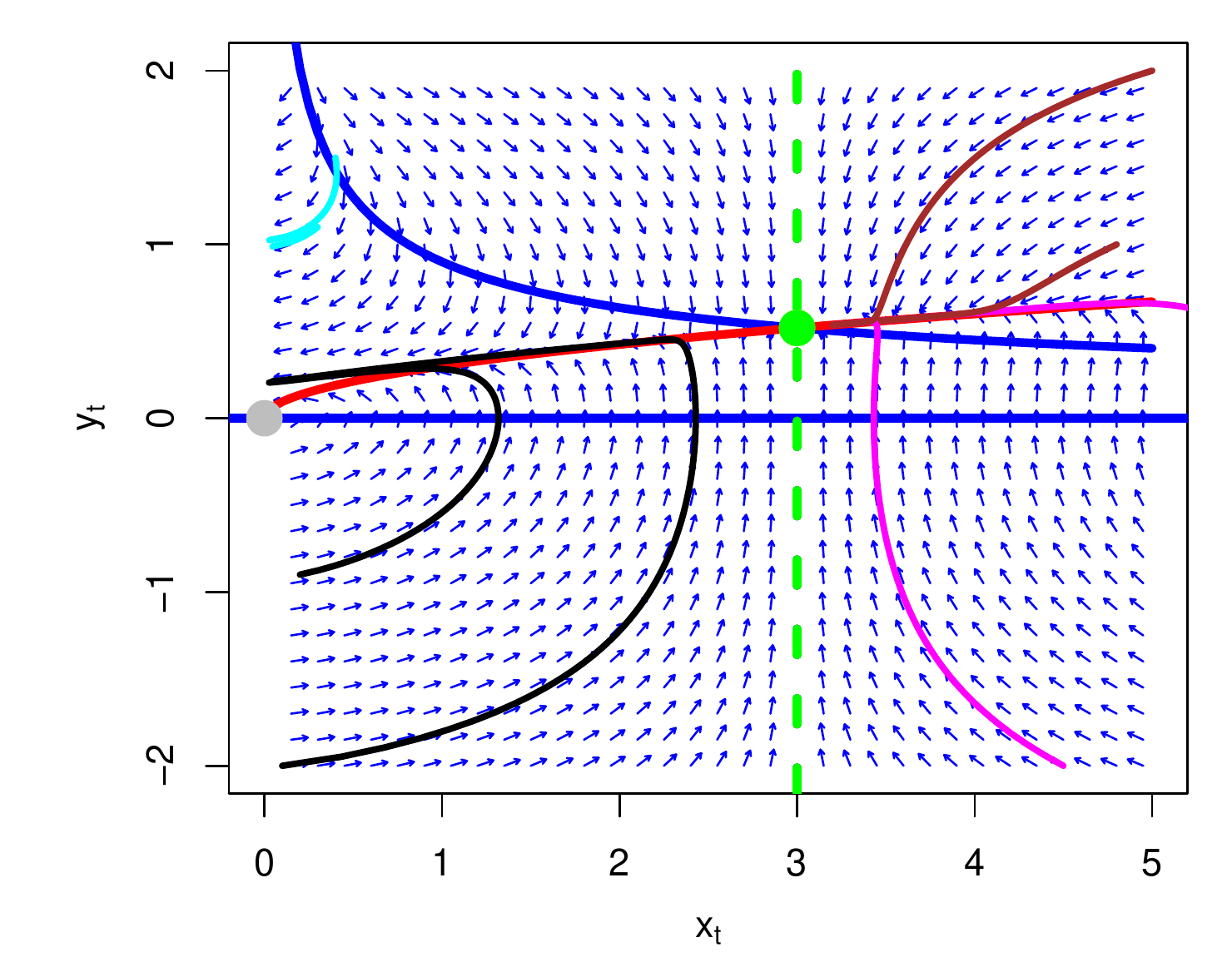}  
  \end{tabular}
   \caption{Time series and phase space diagram when $A=K=3$. The point $(A, u\sqrt{A})=(K, u\sqrt{K})$ is a saddle.}
\label{fig:PSD3}
  \end{figure}
  
    \begin{figure}[H] 
  \centering
  \begin{tabular}{cc}
  \includegraphics[scale=.55]{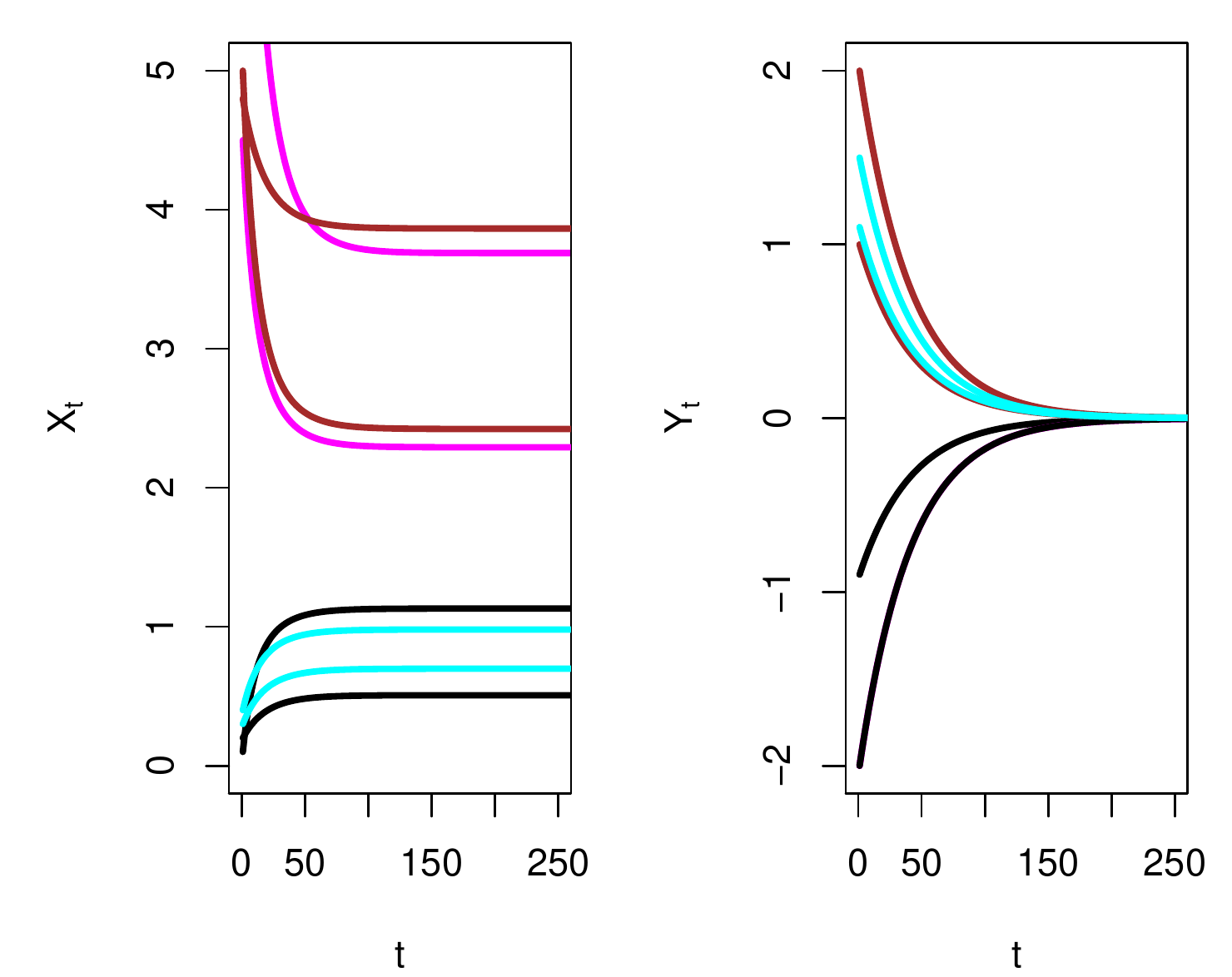}  &  \includegraphics[scale=.55]{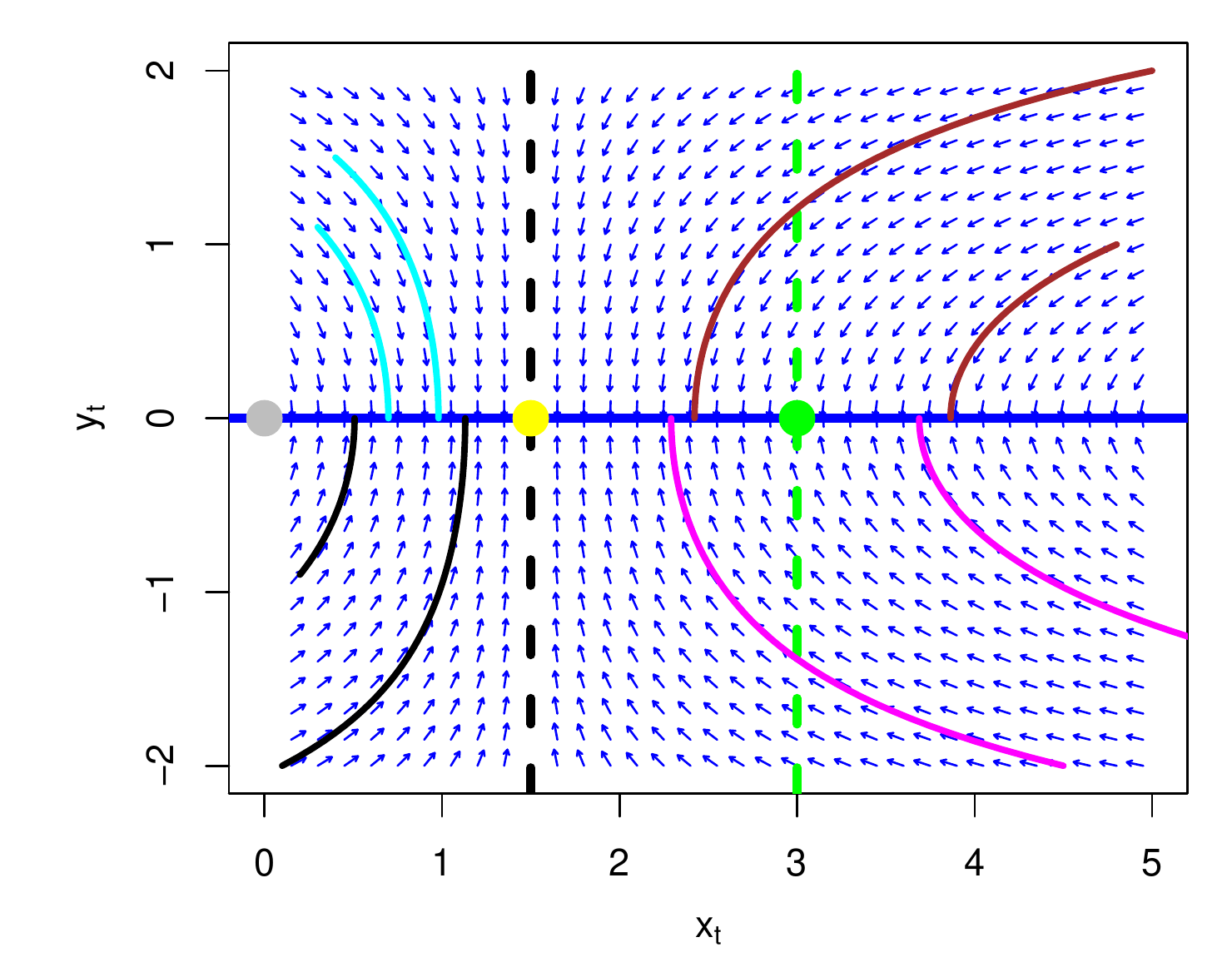}  
  \end{tabular}
   \caption{Time series and phase space diagram when $A=1.5, K=3$,  and  $u=0$.}
    \label{fig:PSD4}
  \end{figure}
      \begin{figure}[H]
  \centering
  \begin{tabular}{cc}
  \includegraphics[scale=.55]{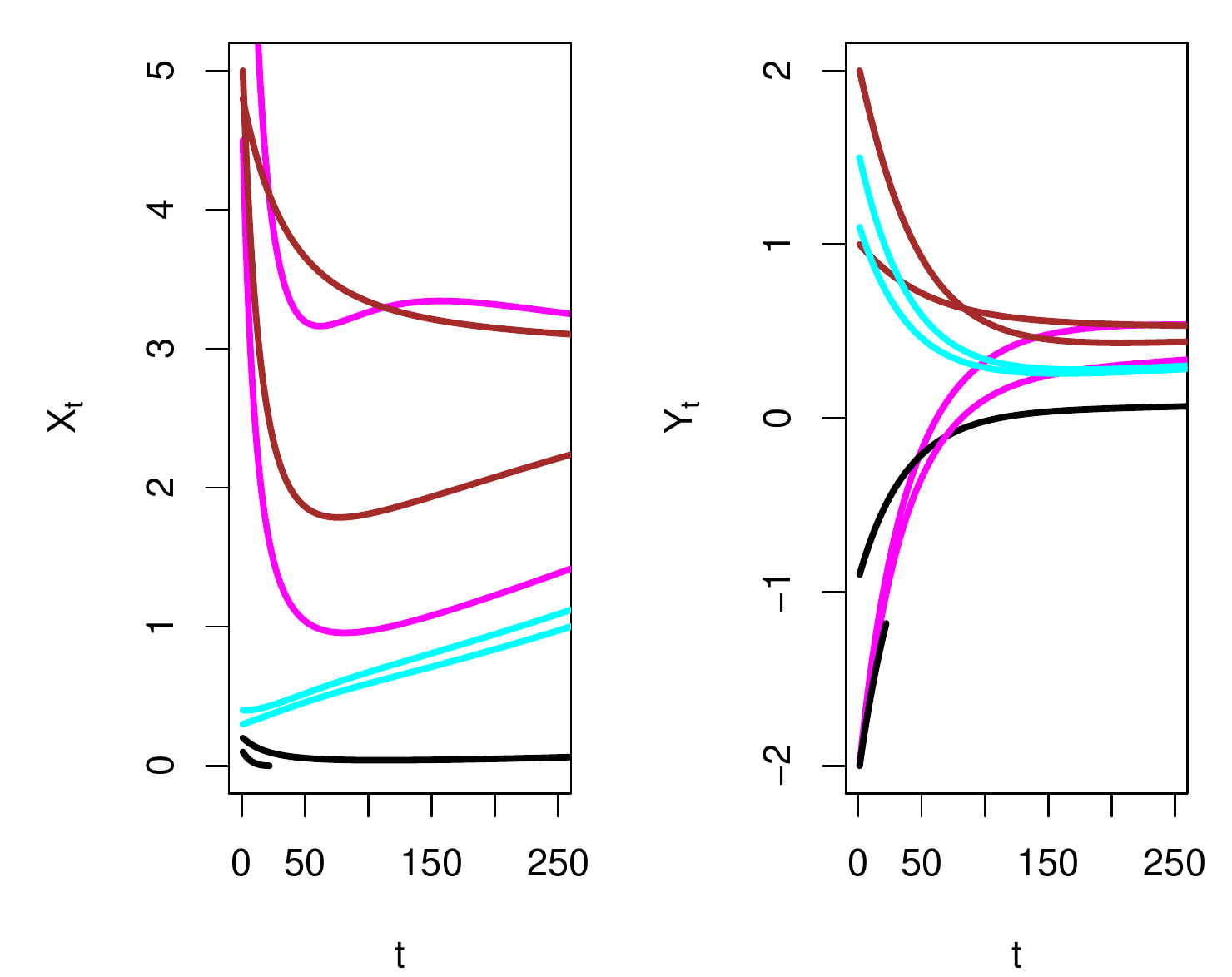}  &  \includegraphics[scale=.55]{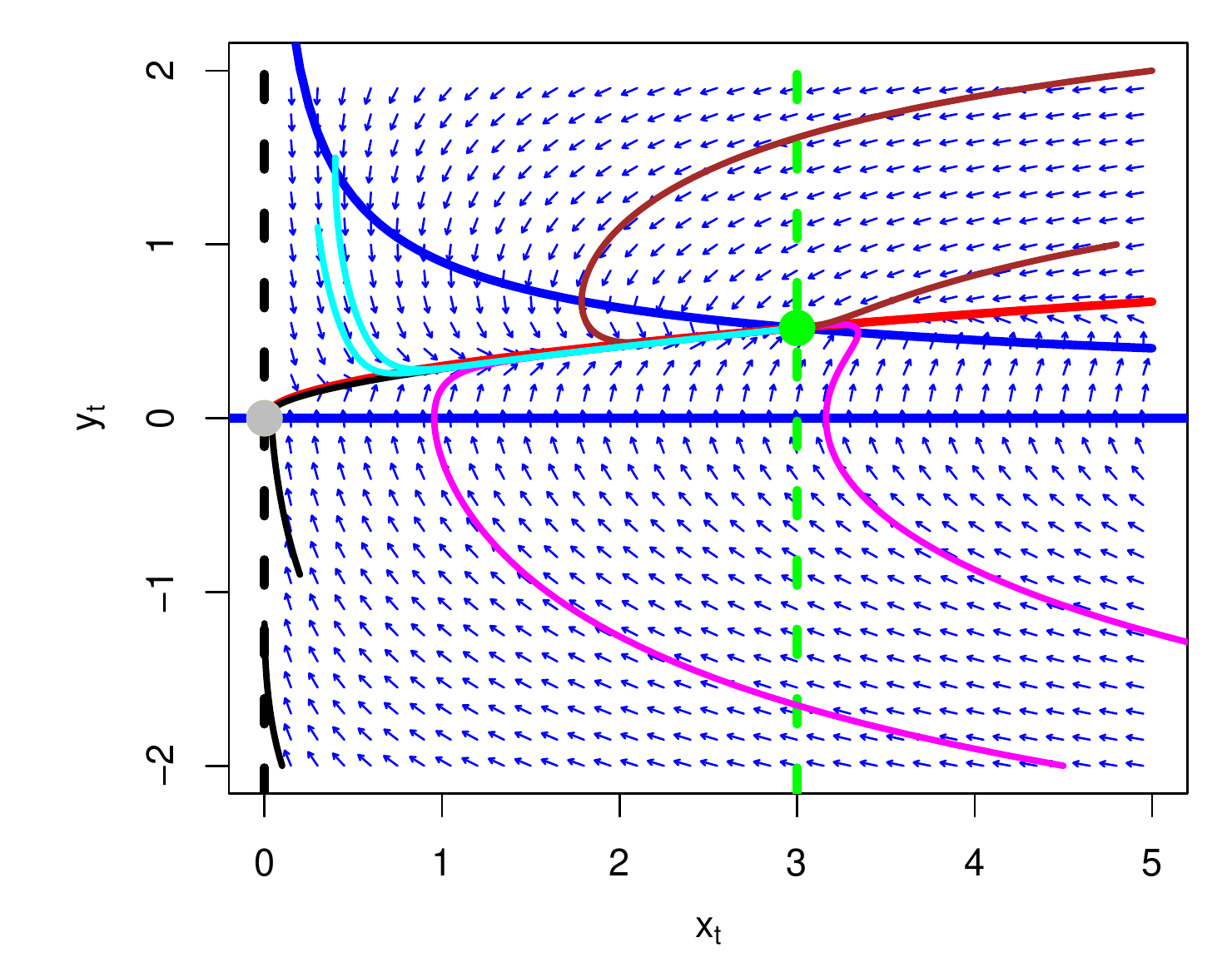}  
  \end{tabular}
   \caption{Time series and phase space diagram when $A=0$ and $K=3$. The point $(K, u\sqrt{K})$ is an attractor as in  the Oja's rule. }
     \label{fig:PSD5}
  \end{figure}
    \subsection{Discussion}
  In the case of one neuron with  non plastic recurrent connection, establishing stability of the steady states  is relatively manageable compared to the case of a  plastic connection. Figure  \ref{fig:PSD1} is similar to a typical case in ecology where the Allee threshold $A$ is less than $K$. We clearly see that in the region below $A$, synaptic weights trajectories converge to zero, while postsynaptic neurons trajectories may be either in  excitatory or inhibitory states. Interestingly, excitatory postsynaptic neurons never become inhibitory since they never cross the red curve. In that same region, inhibitory neurons become excitatory overtime but with decreasing synaptic weights.  Figure  \ref{fig:PSD2} shows that if $A>K$, the Allee effect is no more guaranteed to occur below $A$ or even below $K$. In fact, some neurons, whether in excitatory or inhibitory states would have decaying  or increasing synaptic weights. This is the case for the trajectories in black and cyan color. Figure  \ref{fig:PSD3} is the case when $A=K$ and there is an Allee effect. Figure  \ref{fig:PSD4} is an illustration of  the   situation where at some point in time, there is no presynaptic activity. Since postsynapatic where already in either excitatory or inhibitory modes, they will  decay to zero rather quickly. Figure  \ref{fig:PSD5} is essentially the Oja rule and there is no Allee effect. 
\subsection{Single postsynaptic neuron with one constant  recurrent connection}\label{sect43}

In this case, ${\bf Z}=z$ and ${\bf v}=v$ are constant  with $\ds \frac{d {\bf Z}}{dt}=0$. As we observe  above, we consider the system given by 
\begin{equation}\label{eqn:Alleerule3}
\begin{cases}
\tau_{_{\bf W}}\frac{d\norm{\bf W}^2}{dt}&=2v\left({\bf W}^T\cdot {\bf u}-\frac{v}{K}\norm{\bf W}^2\right)\left(1-\frac{A}{\norm{\bf W}^2}\right)\; \vspace{0.25cm}\\
\ds \tau_{v} \frac{dv}{dt}&=-v+T_2({\bf W, u},z,v)
\end{cases}\;.
\end{equation}
Using the notation $x=:\norm{\bf W}^2,  y:=v, u=\norm{\bf u}\cos(\theta)$,  and  given functions $f_1(x,u)$ and $f_2(x,y,u,v)$,  this system is of the form
\begin{equation}\label{eqn:Alleerule4}
 \begin{cases}
\ds \tau_{x}\frac{dx}{dt} \vspace{0.2cm}&=g_1(x,y):=2y\left(f_1(x,u)-\frac{yx}{K}\right)\left(1-\frac{A}{x}\right)\;\\ 
\ds \tau_y \frac{dy}{dt}&=g_2(x,y):=-y+f_2(x,y,u,z)
\end{cases}\;.
\end{equation}
This system has similar dynamics to that of the system \eqref{eqn:Alleerule22}. In the linear case where $f_2(x,y,u,z)=u\sqrt{x}+zy$, $\ds \frac{dy}{dt}=(z-1)y+u\sqrt{x}=0$ if $(1-z)y=u\sqrt{x}$.  For $z=0$, this is the parabola (red solid curve) obtained in the previous case. When $z$ approaches 1, this parabola becomes increasingly ``steeped" and eventually explodes into the $y$-axis when $z=1$. In the latter  case, there is no steady state in the system since they are always given as intersections between the parabola and the vertical lines $x=A$ and $x=(1-z)K$. In reality, there will be infinitely many points of intersection between the parabola and the vertical lines.

\subsection{Single postsynaptic neuron with one plastic  recurrent connection}\label{sect44}
For a single postsynaptic neuron with one plastic recurrent  connection, we will have  ${\bf Z}=z$ and ${\bf v}=v$.   In this paper, we will use  Goodall's  rule for its de-correlation properties. For a single neuron, we consider the given by 
\begin{equation}\label{eqn:Alleerule44}
\begin{cases}
\tau_{_{\bf W}}\frac{d\norm{\bf W}^2}{dt}&=2v\left({\bf W}^T\cdot {\bf u}-\frac{v}{K}\norm{\bf W}^2\right)\left(1-\frac{A}{\norm{\bf W}^2}\right)\; \vspace{0.25cm}\\
\ds \tau_{v} \frac{dv}{dt}&=-v+T_2({\bf W, u},z,v)\vspace{0.25cm}\\
\ds \tau_{z} \frac{dz}{dt}&= -({\bf W}^T\cdot {\bf u})v+1-z
\end{cases}\;.
\end{equation}
\noindent Let us now discuss the steady states of the system above: \mbox{}\\
{\bf Case 1: $v=0$}\mbox{}\\
Then,  we would have $T_2({\bf W, u},z,v)=0$, which as above, can only happen if $u=0$ and $v=0$. In this case, the third equation suggests that we must have $z=1$. Thus, in the space formed by  $x=\norm{\bf W}^2, y=v$,  and $z$,  the line parallel to the $x$-axis  with equation  $y=0, z=1$  is a steady state.

\noindent {\bf Case 2: $v\neq 0,  {\bf W}^T\cdot {\bf u}=\frac{v}{K}\norm{\bf W}^2$}\mbox{}\\
This condition is equivalent to  $\ds v=\frac{K({\bf W}^T\cdot {\bf u})}{\norm{\bf W}^2}$.\\
From the second equation in \eqref{eqn:Alleerule44}, we have $v={\bf W}^T\cdot {\bf u}+zv$, therefore, we deduce that $z=1-\frac{\norm{\bf W}^2}{K}$.\\ Using the third equation  $-({\bf W}^T\cdot {\bf u})v+1-z=0$, it follows that $\ds z=1-\frac{\norm{\bf W}^2}{K}v^2$. Since the values of $z$ must be the same value, it follows that we should have $v^2=1$. The latter  entails having $\ds \norm{\bf W}^2=K\abs{{\bf W}^T\cdot {\bf u}}$ and $z=1-\abs{{\bf W}^T\cdot {\bf u}}$. We conclude that there are two steady states in the space formed by $x=\norm{\bf W}^2, y=v$,  and $z$, namely, the lines 
\begin{eqnarray*}
L_1: \ds z&=&1-\frac{x}{K},\quad y=1,\\
L_2:  \ds z&=&1-\frac{x}{K}, \quad y=-1\;.
\end{eqnarray*}
{\bf Case 3: $v\neq 0,  \norm{\bf W}^2=A$}\mbox{}\\
In this case,  $v={\bf W}^T\cdot {\bf u}+zv$ and thus $\ds v=\frac{{\bf W}^T\cdot {\bf u}}{1-z}$. We note from above that  if $v\neq 0$, then $z\neq 1$. It follows from the third equation of the system \eqref{eqn:Alleerule44} that, $z=1-({\bf W}^T\cdot {\bf u})v$, and therefore, we can deduce that $(1-z)(1-v^2)=0$. Since $z\neq 1$, we must have $v^2=1$. The latter implies that  $\abs{{\bf W}^T\cdot {\bf u}}=\abs{1-z}$. Thus $z=1\pm {\bf W}^T\cdot {\bf u}$. We conclude that there are two steady states in the space formed by $x=\norm{\bf W}^2, y=v$,  and $z$,  namely,  the points
\begin{eqnarray*}
B_1&=& (A, -1,1+u\sqrt{A}), \quad B_2=(A, 1,1-u\sqrt{A})\;.
\end{eqnarray*}
      \begin{figure}[H]s
\centering \includegraphics[scale=.7]{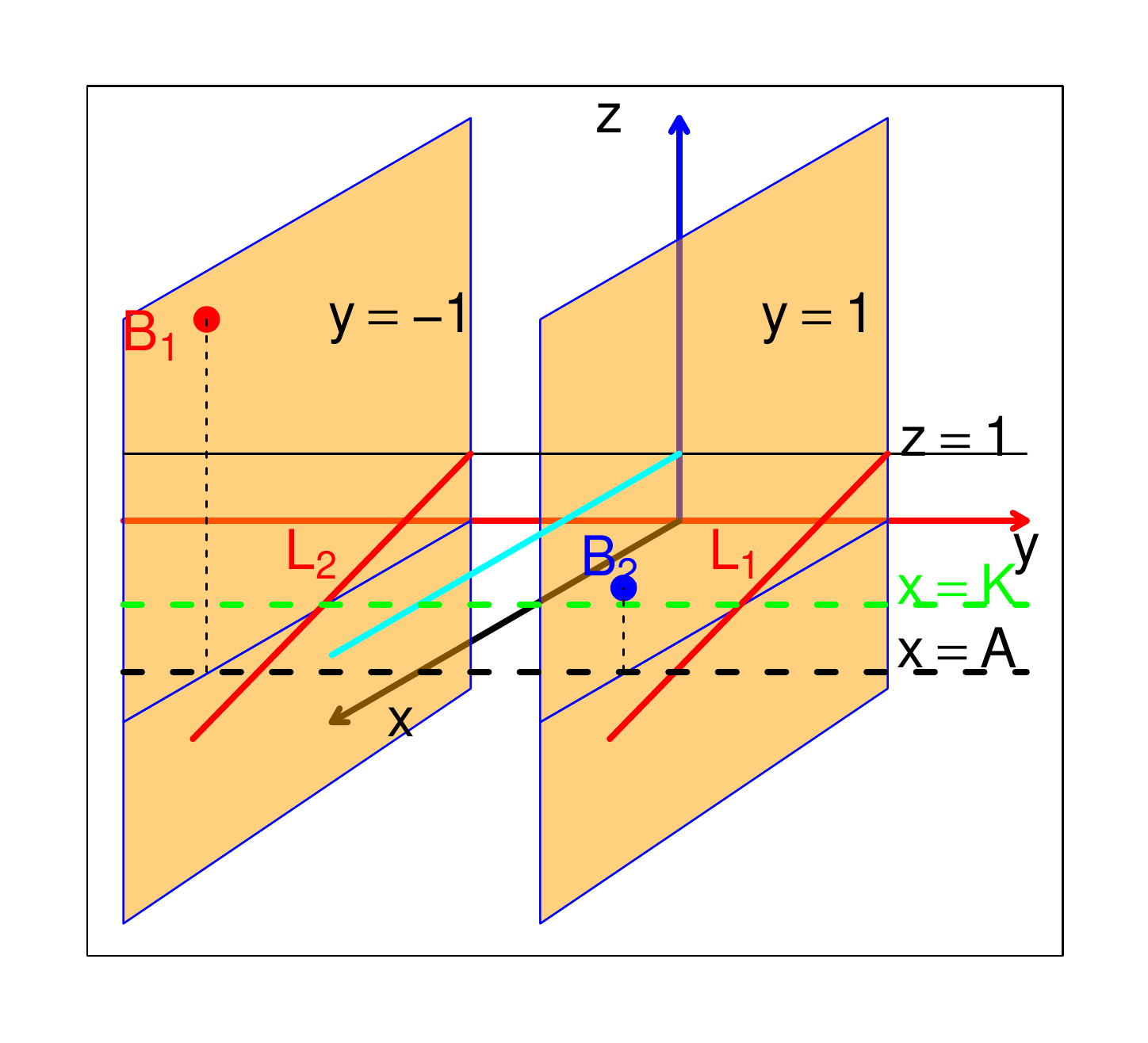}
\caption{Steady states  (in red) of system \eqref{eqn:Alleerule44},  when $A>K$.}
\label{fig8}
\end{figure}

Using the notation $x=:\norm{\bf W}^2, y:=v, u=\norm{\bf u}\cos(\theta) $,  and for given functions $f_1(x,u)$ and $f_2(x,y,z,u)$,  the system \eqref{eqn:Alleerule44} is of the form
\begin{equation}\label{eqn:Alleerule445}
 \begin{cases}
\ds \tau_{x}\frac{dx}{dt} \vspace{0.2cm}&=g_1(x,y,z):=2y\left(f_1(x,u)-\frac{yx}{K}\right)\left(1-\frac{A}{x}\right)\;\\ 
\ds \tau_y \frac{dy}{dt}&=g_2(x,y,z):=-y+f_2(x,y,z,u)\\
\ds \tau_{z} \frac{dz}{dt}&=g_3(x,y,z):= f_3(x,y,u)+1-z
\end{cases}\;.
\end{equation}
Our first result in this section concerns the stability of the steady state line $x>0, y=0,z=1$ and the points $B_1$
 and $B_2$.
 \begin{thm} \label{thm220} Consider the system \eqref{eqn:Alleerule445}, where $f_1(x,u)=u\sqrt{x}, f_2(x,y,z,u)=u\sqrt{x}+zy$, and $f_3(x,y,u)=-uy\sqrt{x}$. 
\begin{itemize}
\item[(i)] If $u=0$, then the steady state is the line $(x,0,1)$ for $x>0$ and it is always stable.
\item[(ii)] Suppose $u>0$.
\begin{itemize}
 \item[(a)] If $\ds u<\frac{2\sqrt{A}}{K}$, then $B_1$ is unstable and $B_2$ is stable.

 \item[(b)] If $\ds u>\frac{2\sqrt{A}}{K}$, then $B_1$ and $B_2$ are unstable.
 \end{itemize}
 \item[(iii)] Suppose $u<0$.
 \begin{itemize}
 \item[(a)] If $\ds -2\sqrt{A}\max\set{2A, K}\leq u$, then $B_1$  and $B_2$ are  stable.

 \item[(b)] If $\ds u<-2\sqrt{A}\max\set{2A, K}$, then $B_1$ and $B_2$ are unstable.
  \item[(c)] If $\ds -2\sqrt{A}\min\set{2A, K}< u<-2\sqrt{A}\max\set{2A, K}$, one of  $B_1$ or $B_2$ is  unstable and the other is stable.
 \end{itemize}
\end{itemize}
\end{thm}
The proof can be found in the Appendix.  
\noindent Our second result discusses the stability of the  steady states $L_1$ and $L_2$.
\begin{thm} \label{thm221} 
Consider the system  \eqref{eqn:Alleerule445}, where $f_1(x,u)=u\sqrt{x}, f_2(x,y,z,u)=u\sqrt{x}+zy$, and $f_3(x,y,u)=-uy\sqrt{x}$. Put
\begin{eqnarray*}
   \alpha_1&=&2\rb{\frac{u}{2\sqrt{x}}-\frac{1}{K}}\rb{1-\frac{1}{A}}+2\rb{u\sqrt{x}-\frac{2x}{K}}\rb{\frac{A}{x^2}}, \\
   \alpha_2&=& \ds\rb{ 2u\sqrt{x}-\frac{4x}{K}} \left(1-\frac{A}{x}\right),\quad \quad \alpha_3=0\;,\\
   \ds \beta_1&=&\frac{u}{2\sqrt{x}},\quad \quad \beta_2= -\frac{x}{K}, \quad \quad \beta_3=-1\;,\\
   \ds \gamma_1&=&\frac{u}{2\sqrt{x}}, \quad \quad \gamma_2=-u\sqrt{x}, \quad \quad \gamma_3=-1\;.
   \end{eqnarray*}
   In addition, we let 
   \begin{eqnarray*}
   a_2&=&(\alpha_1+\beta_2-1)\;,\\
   a_1&=&\alpha_1(1-\beta_2)+\beta_2-\gamma_2+\alpha_2\beta_1\;,\\
   a_0&=& \alpha_1(\gamma_2-\beta_2)\;.
   \end{eqnarray*}
\begin{itemize}
\item[(i)] Suppose $a_0=0$.
\begin{itemize}
\item[(a)] If $a_2^2+4a_1<0$ and $a_2<0$, then $L_1$ and $L_2$ are  stable.
\item[(b)] Suppose $a_2^2+4a_1>0$.
\begin{enumerate} 
\item If $a_1>0$ and $a_2>0$,  then $L_1$ and $L_2$ are  unstable.
\item If $a_1<0$ and $a_2>0$, then $L_1$ and $L_2$ are stable.
\item If $a_1>0$ and  $a_2<0$ or $a_1<0$ and  $a_2>0$,   then $L_1$ and $L_2$ are unstable.
\end{enumerate}
\end{itemize}
\item[(ii)] If  $a_0<0$, then  $L_1$ and $L_2$ are unstable in two cases and stable in one.
\item[(iii)] If  $a_0>0$, then  $L_1$ and $L_2$ are unstable in case  and stable two cases.
\end{itemize}
\end{thm}
The proof can be found in the Appendix 
\subsubsection*{Illustration}
Since they are. any cases to consider, we will consider for simplicity just a couple of them for sake of simplicity. In the figures below, we show the dynamics of the system \eqref{eqn:Alleerule44} above. We choose $M=20$ different trajectories with length $N=5000$. The initial state are chosen randomly as: since $x=\norm{\bf W}^2$ must be positive, we randomly select $M$ starting points  $x_0$ in the interval $(0,5)$.  The starting values $y_0$ are chosen randomly as $M/2=10$ in the interval $(-5,0)$ and $M/2=10$ in $(0,5)$. The starting values $z_0$ are chosen randomly as $M/2=10$ in the interval $(-10,0)$ and $M/2=10$ in $(0,10)$. The starting points $(x_0,y_0,z_0)$ are the white dots in the figures below. The green sphere is $B_1$ while the blue sphere is $B_2$.

\begin{figure}[H] 
   \centering
   \includegraphics[scale=.5]{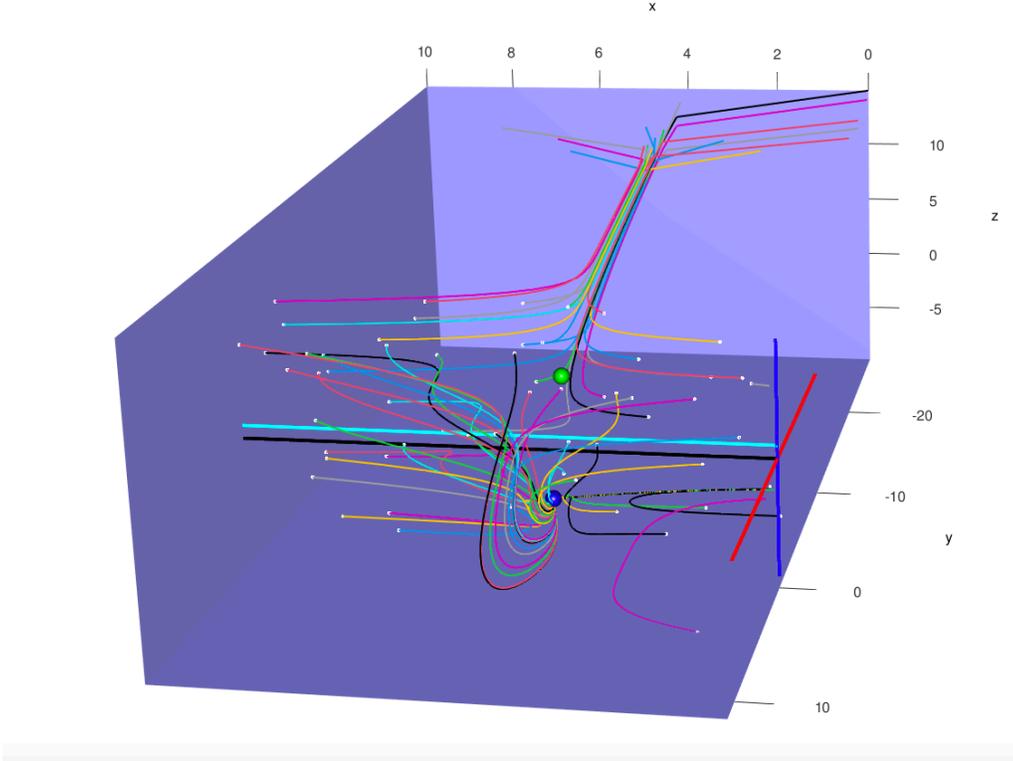} 
   \caption{Illustration of the dynamics of the system above for $u=2,K=1$, and $A=4$. $u<\frac{2\sqrt{A}}{K}$ and we observe that $B_2$ is stable and $B_1$ is unstable. The cubes $R_1=[0,4]\times [0,10]\times[-5,0]$ and $R_2=[0,4]\times [-20,0]\times[-5,10]$   are  the Allee regions: starting trajectories will eventually converge to 0 in $x$ leading to absence of plasticity.}
   \label{fig:example}
\end{figure}

\begin{figure}[H] 
   \centering
   \includegraphics[scale=0.5]{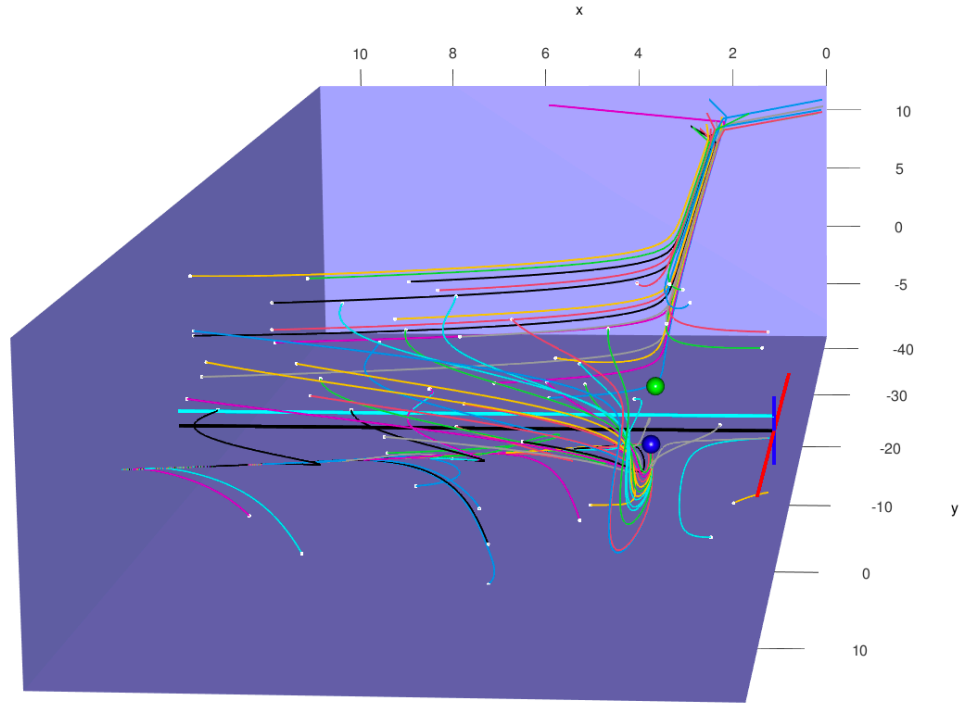} 
   \caption{Illustration of the dynamics of the system above for $u=1.1,K=3$, and $A=2$. $u>\frac{2\sqrt{A}}{K}$ and we observe that $B_1$  and $B_2$ are  unstable. In fact, $B_1$ is clearly a repeller point whereas $B_1$ is a saddle point.  This case is less realistic since synaptic normalization is never achieved. In fact,  the lengths of weights $x$ increase without bound while the postsynaptic activities $y$, though different at early times, become increasing similar overtime (straight line). The Allee regions are the cubes $R_1=[0,2]\times[0,10]\times[-5,0]$ and $R_2=[0,2]\times[-40,0]\times[-5,10]$. We observe that the gray trajectory starting just above $R_1$ does not converge to 0 because it gets into the basin of attraction of $L_2$ and increases thereafter. }
   \label{fig:example}
\end{figure}

\begin{figure}[H] 
   \centering

   \includegraphics[scale=.5]{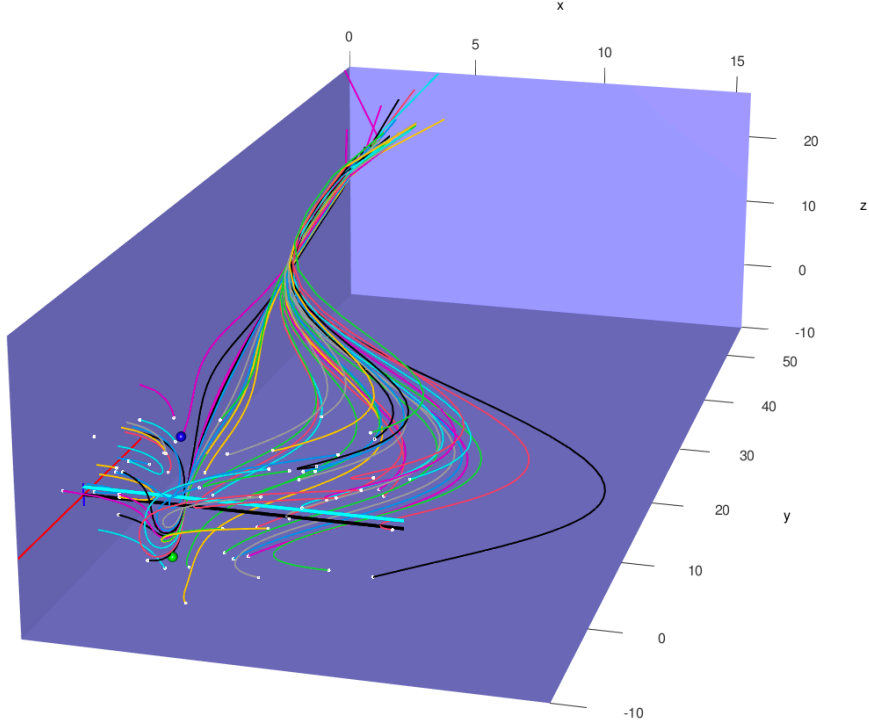}
   \caption{Illustration of the dynamics of the system above for $u=-4,K=4$, and $A=4$. $u>-2\sqrt{A}\max\{2A,K\}$ and we observe that $B_1$  and $B_2$ are  unstable.  This case is less realistic since synaptic normalization is not achieved all the time. Moreover,  postsynaptic activities $y$ and recurrent connections $z$ increase without any bound. If we look close to $B_1$ and $B_2$, we  clearly see that no trajectory converges to $B_1$ and $B_2$. The Allee region is  cube $R_1=[0,4]\times [0,50]\times[-10,20]$. }
   \label{fig:example}
\end{figure}





  \subsection{Discussion}
The first simulation shows that lengths of weights $x$ either decrease to $A=4$ or to 0. The second simulation is more nuanced in that some  decrease to $A=2$ first, then after a while they either decrease to 0 or increase. Others will first increase and  then decrease to 0. Finally, some will increase without bound after initially decreasing to close to $A=2$.
What these simulations show is that the size of the  Allee regions depend on the value of $A$. Clearly, if $A=0$, there is no Allee region and  the model is reduced to the Oja  rule.  An important observation is that since the fixed points $B_1$ and $B_2$ both depend on $A$, the first simulation shows that if $A=0$ (Oja rule), then weight lengths $x$ all decrease to 0, without any possibility of recovering. This means that, our system while stable in the longterm, represents adrift  towards an absence of plasticity. In a sense, the parameter $A$ must positive if we want to have more than a drift towards absence of plasticity for all trajectories.
 
From above, we clearly see that there is advantage studying the length of weights rather than individual weights. The complexity of the dynamics is vastly reduced. This approach makes studying large amounts of layers mathematically possible while maintaining interpretability of the results. The main drawback, as illustrated by the results above (Theorem \ref{thm220}  and  \ref{thm221}) is that stability analysis, while feasible, still depends unfortunately of complicated quantities. 
  
 \section{Multiple postsynaptic neurons model}\label{sect6}
In the single postsynaptic neuron model, we did not include recurrent connections. For a multiple postsynaptic model, we have to consider recurrent connections which themselves maybe fixed  or plastic. Another important aspect to consider is that of a layered system where pre-and postsynaptic neurons are on different layers. It is entirely possible that this aspect may help reduce  redundancies and correlations among output units.
\subsection{Multiple output units with constant recurrent connections}\label{sect66}
In this case, ${\bf Z}$ and ${\bf v}$ are nonzero matrices with $\bf Z$ constant over time. In this case, we fix $1\leq \ell\leq L$ and $1\leq j\leq N_v$. The assumption here is still that we have $N_u$ presynaptic neurons and $N_v$ post-synaptic neurons per layer. Now we let ${\bf W}_j^{(\ell)}$ be $1\times N_u$  vector of synaptic weights from the $N_u$ presynaptic neurons ${\bf u}^{(\ell)}$ on the $\ell$th layer to the  $j$th postsynaptic neurons $v_j^{(\ell)}$ on the $\ell$th layer.  In this case, equation \eqref{eqn:Alleerule3}

\begin{equation}\label{eqn:Alleerule34}
\begin{cases}
\ds \tau_{_{{\bf W}_j^{(\ell)}}}\frac{d\norm{{\bf W}_j^{(\ell)}}^2}{dt}&=\ds 2v_j^{(\ell)}\left([{{\bf W}_j^{(\ell)}}]^T\cdot {\bf u}^{(\ell)}-\frac{v_j^{(\ell)}}{K_j^{(\ell)}}\norm{{\bf W}_j^{(\ell)}}^2\right)\left(1-\frac{A_j^{(\ell)}}{\norm{{\bf W}_j^{(\ell)}}^2}\right)\;\\
\ds \tau_{v_j^{(\ell)}} \frac{dv_j^{(\ell)}}{dt}&=-v_j^{(\ell)}+T\left({\bf W}_j^{(\ell)}, {\bf u}^{(\ell)},z_j^{(\ell)},v_j^{(\ell)}\right)
\end{cases}\;.
\end{equation}
We see that for these given $\ell$ and $j$, the system \eqref{eqn:Alleerule34} has similar dynamics as the system is \eqref{eqn:Alleerule3}. However, there are  other considerations to account  for in this case. The system's parameters all depend on $\ell$ and $j$. The assumptions that  time scale constants $\tau_{_{{\bf W}_j^{(\ell)}}}$ are the same is not completely unrealistic, especially if the system evolves in a homogeneous ambient space. The same can be said of time scale constants $\tau_{_{ v_j^{(\ell)}}}$. The thresholds $A^{(\ell)}$ and $K^{(\ell)}$ can either be  the same or can vary, selected according to a chosen distribution. In the constant case, the dynamics of the system \eqref{eqn:Alleerule34} is identical across all layers, thus the postsynaptic neurons will be perfectly correlated.  In the case where these thresholds are not identical, of interest is understanding how and if the thresholds vectors  $\bm{A}^{(\ell)}=(A_j^{(\ell)})$ and $\bm{K}^{(\ell)}=(K_j^{(\ell)})$ for $ 1\leq j\leq N_v, 1\leq \ell \leq L$ affect the correlation between postsynaptic neurons ${\bf v}^{(\ell)}$ per layer. 

\begin{figure}[H]
  \centering
  \begin{tabular}{cc}
  (a) & (b)\\
  \includegraphics[scale=.58]{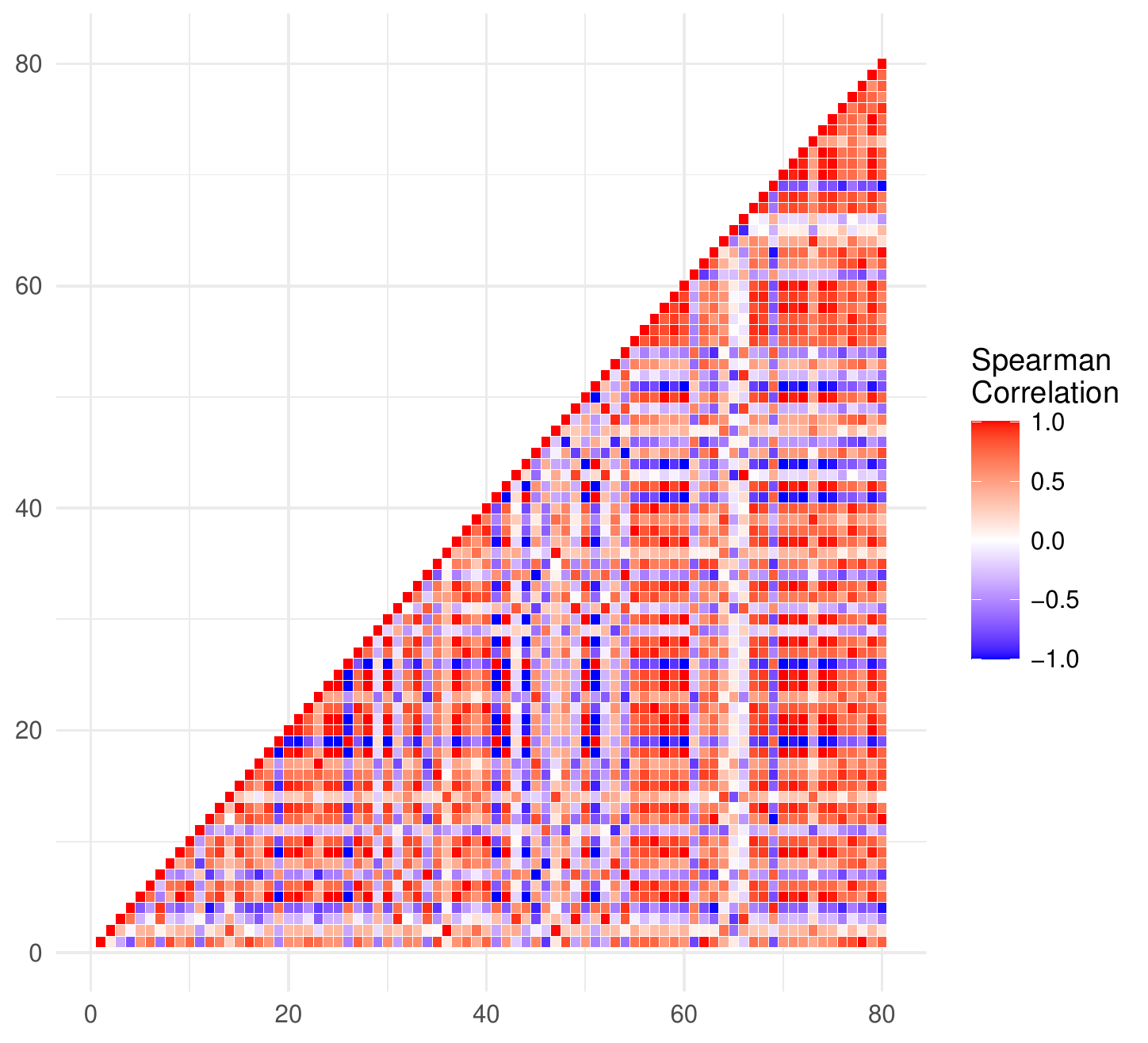}  &  \includegraphics[scale=.45]{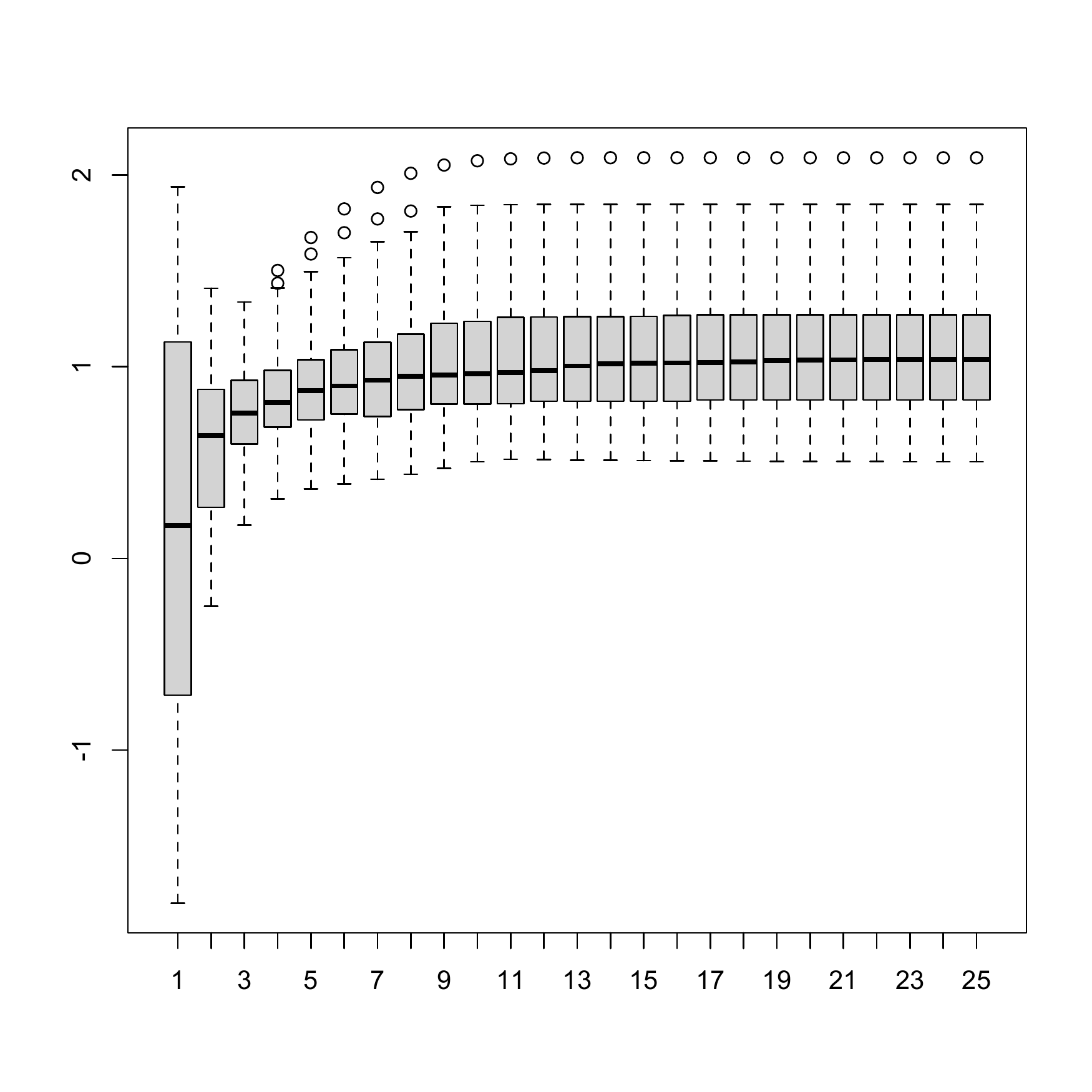}  
  \end{tabular}
   \caption{(a) Heatmap showing correlation between the $N_v^*=80$ convergent trajectories. (b) Boxplots showing the evolution of the distribution of trajectories form time $t=1$ to $t=25$.}
     \label{fig:Heatmap1}
  \end{figure}

To illustrate the potential effect of thresholds, we will select select  $N_v=150$ samples of  $K$ from a truncated normal distribution $N(\mu=0, \sigma^2=100)$ over an interval $[1.5,30]$. Likewise, we will select $N_v$ samples $A$ from an exponential distribution $\exp(\theta=0.5)$. These distributions are different enough to discriminate potential effect of thresholds $A$ and $K$. The synaptic weights lengths $x=\norm{\bf W}^2$ will be initialized uniformly over the interval $(0,5)$. We fix the presynaptic length $u=0.3$ and we let $z_j^{(\ell)}=0.4$. The postsynaptic values $v$ will be initialized uniformly within the interval $[-2,2]$.  We will observe the $N_v$ trajectories of $v$ from $t=0$ to $t=25$, because not all of them will converge. In fact,  for given $N_v$ postsynaptic neurons, only a $N_v^*\leq N_v$ will converge. We will therefore assess the correlation between these $N_v^*$ trajectories. In  Figure \ref{fig:Heatmap1} (a) below, the heat-map shows that the majority of the $N_v^*=80$  postsynaptic neurons $v$ are highly correlated. Some of them, albeit a small number are de-correlated. It could be due to the randomness in the choice of the parameters above, or it could be due to the fact that the models itself reduces correlation, without any formal de-correlation  mechanism  like Goodall's. The boxplots in Figure \ref{fig:Heatmap1} (b) show the evolution of the $N_v^*$ trajectories overtime. While the distribution of the $N_v^*=80$ trajectories  differ significantly initially,  they become increasingly similar over time, despite a few outliers. It also shows that the variance of outputs is constant overtime. 

\begin{figure}[H]
 \centering 
 \includegraphics[scale=.55]{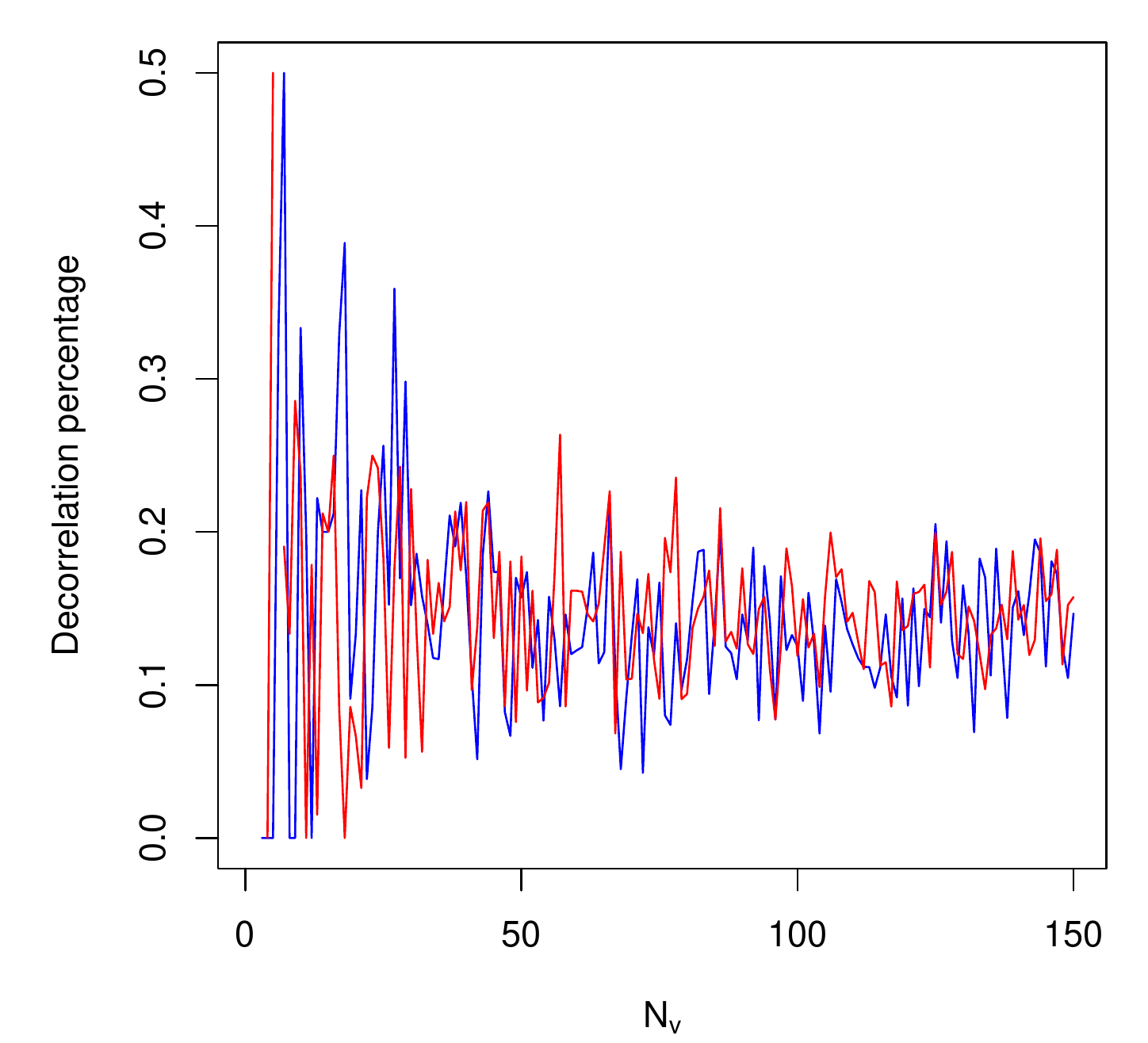}  
   \caption{De-correlation percentage for the Oja model (red) and the Allee model (blue), both as functions of the number of postynaptic neurons per layers $N_v$.}
     \label{fig:Decorrelation}
  \end{figure}

  To ascertain whether the number of de-correlated postsynaptic neuron $v$ is independent of the number $N_v$ chosen, we introduce the de-correlation percentage. There are $\ds {N_v^* \choose 2}$ Spearman correlation coefficients. The   de-correlation  percentage is the proportion of these coefficients less than 0.2 (considered a weak correlation in the literature).  Figure
\ref{fig:Decorrelation} shows that the de-correlation percentage is high when $N_v$ is low, and decrease with increasing  $N_v$. 
\subsection{Multiple output units with plastic recurrent connections}\label{sect67}
In this case, ${\bf Z}$ and ${\bf v}$ are matrices where  $\bf Z$  is time-dependent. As  Section \ref{sect66} above, we will fix $1\leq k, \ell \leq L$ and $1\leq  j,m\leq N_v$. Let  $z_{mj}^{(k,\ell)}$ represents  the plastic weight connecting the $j$th postsynaptic neuron $ v_j^{(\ell)}$ on the $\ell$th layer with the $m$th postsynaptic neuron $ v_m^{(k)}$ on the $k$th layer. The system \eqref{eqn:Alleerule44} becomes:

\begin{equation}\label{eqn:Alleerule444}
\begin{cases}
\ds \tau_{_{{\bf W}_j^{(\ell)}}}\frac{d\norm{{\bf W}_j^{(\ell)}}^2}{dt}&=\ds 2v_j^{(\ell)}\left([{{\bf W}_j^{(\ell)}}]^T\cdot {\bf u}^{(\ell)}-\frac{v_j^{(\ell)}}{K_j^{(\ell)}}\norm{{\bf W}_j^{(\ell)}}^2\right)\left(1-\frac{A_j^{(\ell)}}{\norm{{\bf W}_j^{(\ell)}}^2}\right)\; \vspace{0.1cm}\\
\ds \tau_{v_j^{(\ell)}} \frac{dv_j^{(\ell)}}{dt}&=-v_j^{(\ell)}+T\left({\bf W}_j^{(\ell)}, {\bf u}^{(\ell)},z_{mj}^{(k,\ell)},v_j^{(\ell)}\right)
\vspace{0.25cm}\\
\ds \tau_{_{z_{mj}^{(k,\ell)}}} \frac{d z_{mj}^{(k,\ell)}}{dt}&= -\left([{{\bf W}_j^{(\ell)}}]^T\cdot {\bf u}^{(\ell)}\right)v_j^{(\ell)}+1-z_{mj}^{(k,\ell)}
\end{cases}\;.
\end{equation}
As above, we may assume that the ambient space is homogeneous so that  time scale constants $\tau_{_{z_{mj}^{(k,\ell)}}} $ are the same.  To obtain Figure  \ref{fig:Heatmap2} below, we used the same parameters as in Section \ref{sect6}, with the addition of plastic recurrent connections.  The Heatmap in Figure  \ref{fig:Heatmap2} (a), shows that the $N_v^*=107$ postsynaptic neurons are less correlated than in the previous case above based on the prevalence of light red and light blue colors. Figure  \ref{fig:Heatmap2} (b) shows that the distribution of the trajectories stabilizes relatively quickly compared to the case above. 

 \begin{figure}[H]
  \centering
  \begin{tabular}{cc}
  (a) & (b)\\
  \includegraphics[scale=.58]{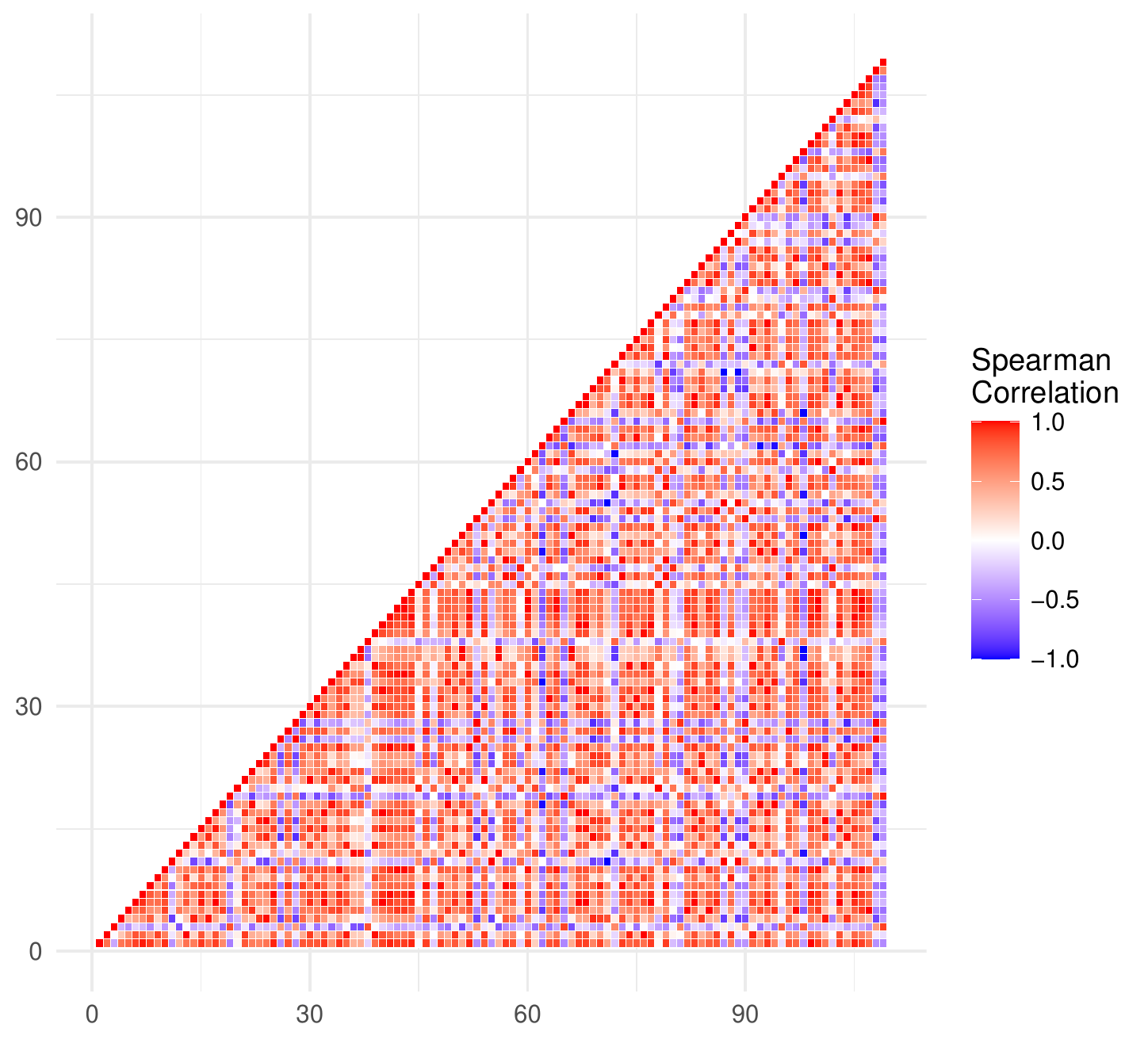}  &  \includegraphics[scale=.45]{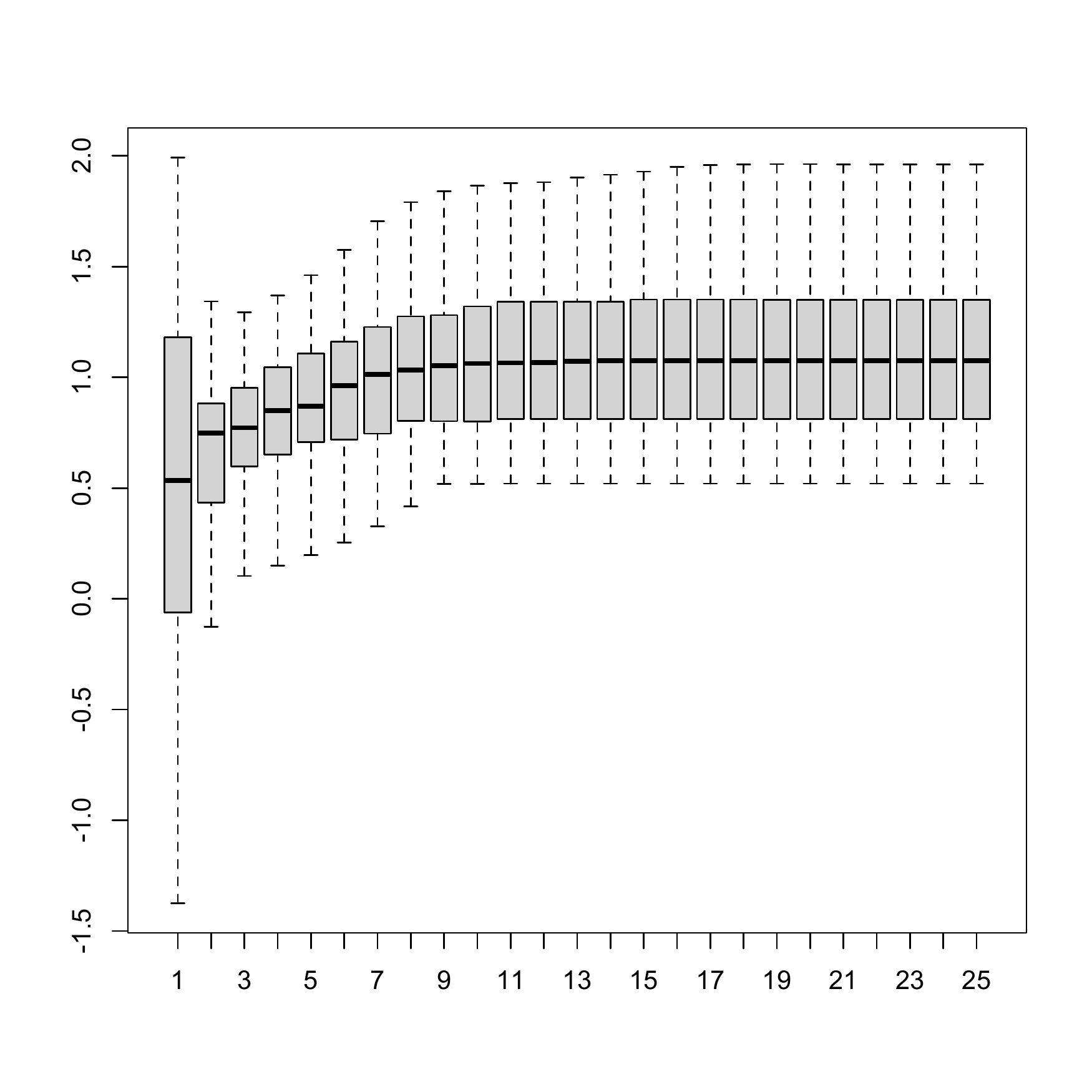}  
  \end{tabular}
   \caption{a) Heatmap showing correlation between the $N_v^*=107$ convergent trajectories. (b) Boxplots showing the evolution of the distribution of trajectories form time $t=1$ to $t=25$. }
     \label{fig:Heatmap2}
  \end{figure}
  
   \begin{figure}[H]
 \centering 
 \includegraphics[scale=.55]{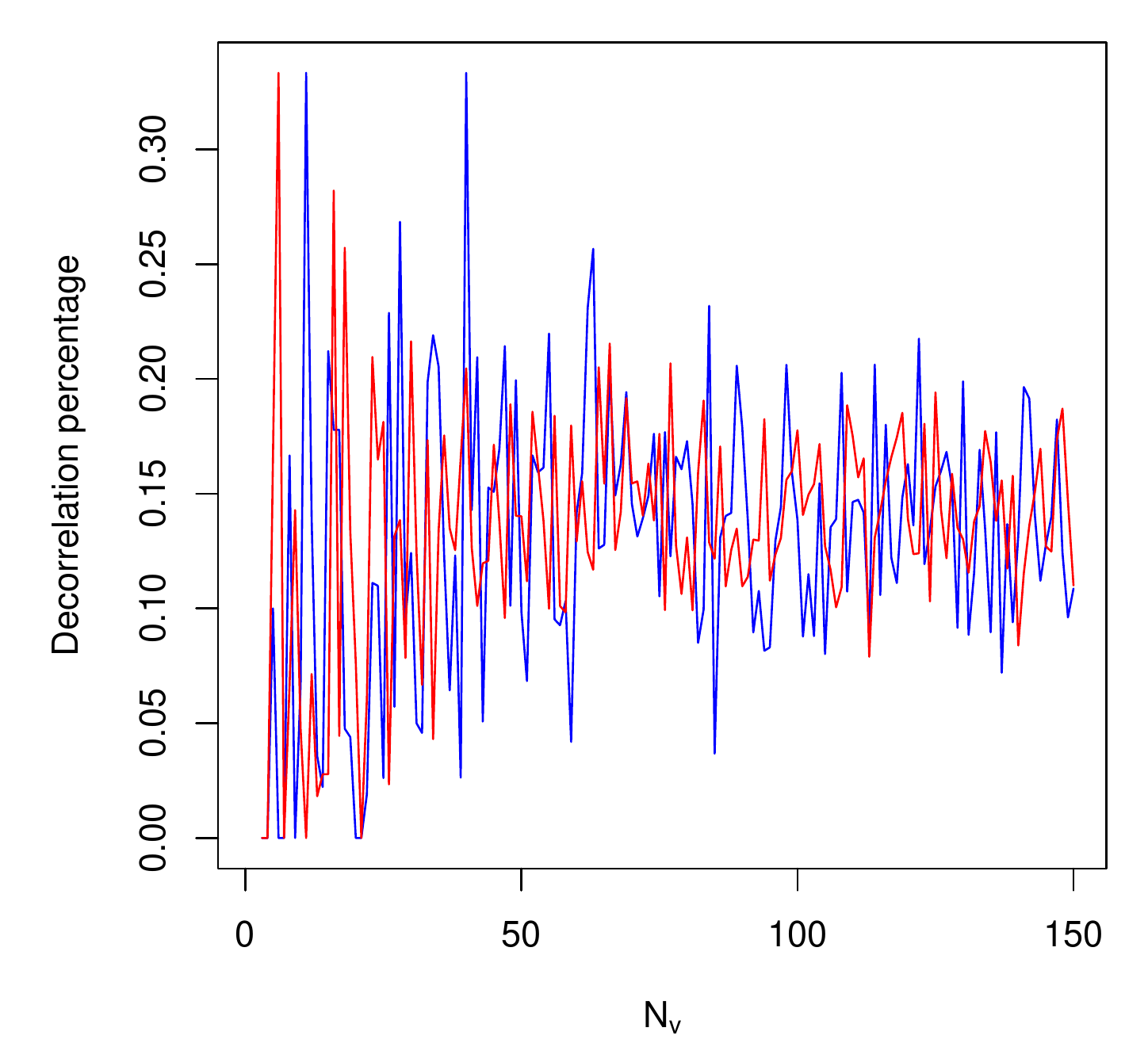}  
   \caption{De-correlation percentage for the Oja model (red) and the Allee model (blue), both as functions of the number of postynaptic neurons per layers $N_v$.}
     \label{fig:Decorrelation1}
  \end{figure}
\subsection{Discussion}
From the simulations above, we can draw a few observations:\\
1) The presence of the Allee effect term $(1-A\norm{\bf W}^{-2})$ in the model overall increases de-correlation, relatively speaking, see Figure \ref{fig:Decorrelation} above. De-correlation is even increased when coupled with a de-correlation mechanism such as the Goodall's method.  \\
2) We selected the thresholds $A$ and $K$ randomly, from noisy distributions so that any measured effects would  be independent of  their selection. Another important observation is that the initialization of $v$ and ${\bf Z}$ does not seems to produce similar results as seen  above, even when choosing from heavy-tailed distributions like $N(0,10)$ or a Student-t with low degrees of freedom.\\
3) The systems \eqref{eqn:Alleerule34} and  \eqref{eqn:Alleerule444} are discussed in the context where the time scales are identical per layer. However, if they are chosen to be different and large, then the output units  become highly correlated, within layers, reversing the de-correlation gains an Allee term would bring. \\
4) It  seems as though the model, as written in \eqref{eqn:Alleerule34} and  \eqref{eqn:Alleerule444} may be local to a single chosen  layer. However,  it is hardly the case given that one can consider that each layer has one single postsynaptic neurons, similarly to discrete dynamics of dynamic neural fields, see \cite{Kwessi2021_5}. 
\section{Conclusion}

In this paper, we have proposed a definition of the Allee effect in neuroplasticity that maintains the spirit of the Allee effect as originally proposed by \cite{Allee1949}. We have also proposed a learning rule  that is more general than the Oja  learning rule while preserving multiplication normalization, controlling unbounded growth and inducing competition between weights. The model has the advantage that it can accommodate single or multiple pre-and postsynaptic neurons, with and and without recurrent connections. Stability analysis  was discussed with  simulations to illustrate results. Absence of plasticity in the brain can be due to many factors and can be observed in many brain pathologies such as the Alzheimers, stroke, Parkinson's and Huntington disease. Using the firing-rate equation to model post-synaptic activities could be a limiting factor in the model in that diffusion is not accounted for. A further improvement involving a diffusion term or  a lattice differential equation would probably add more nuance and  could be worthwhile. 
In Ecology, remedies to an Allee effect such as immigration have been proposed in \cite{Kwessi2015_6}, together with mathematical models  explaining the process. This conservationism  amounts in practice to adding either new  offsprings from different population patches or to  controlling predation.  In the brain however, it is not clear how one would go about this. Often, neuroscientists focus on reactivating lost or dormant neurotransmitters by using new technologies such as brain implants. While applications of these brain implants have been numerous, mathematical models have lagged or at best, they have have  been adaptation or reduction  of the Hogkin-Huxley model, see \cite{Drapaca2018}.  The model  we proposed is different and is sensitive to external inputs (as in an Alllee-type model) and thus could be used to model the effects of brain implants.  From a dynamical systems point of view, the model we propose is complex enough to accommodate a complex structure  like the brain. Simulations do show that there are enough parameters in the model to capture a variety of phenomena related to neuroplasticity. From a mathematical point of view, studying the length of weights (rather than individual weights) coupled with other numerical quantities such as postsynaptic signals and their strengths reduces the complexity of the problem while
maintaining  interpretability of the results. 
From a purely scientific point of view, the model offers to merge notions discussed in ecology and neuroscience and it  shows that these seemingly isolated areas from each other actually have similarities. 
\bibliography{Plasticity}
\section{Appendix}
\subsection{Appendix A: Stability analysis of System \eqref{eqn:Alleerule22}}
\begin{proof}
Per remark \ref{rem1}  above,  we have $y=f_2(x,u)=0$ for all $x>0$ only when $u=0$. Therefore the Jacobian matrix $J(x,0)$ is given as 
\[ 
J(x,0)=\begin{pmatrix} 
0  \vspace{0.2cm} & \ds 2f_1(x,0)\left(1-\frac{A}{x}\right)\\
\ds0 & \ds -1
 \end{pmatrix}\;.\]
 The eigenvalues are therefore $\lambda_1=0$ and $\lambda_2=-1$, which implies that the line $y=0$ is always a spiral sink.\\

The partial derivatives of $g_1$ and $g_2$ are
\begin{eqnarray*}
\frac{\pa g_1(x,y)}{\pa x}&=& 2y\left(\frac{\pa f_1(x,u)}{\pa x}-\frac{y}{K}\right)\left(1-\frac{A}{x}\right)+2y\left( f_1(x,u)-\frac{yx}{K}\right)\left(\frac{A}{x^2}\right),\\
 \frac{\pa g_1(x,y)}{\pa y}&=& \left( 2f_1(x,u)-4\frac{yx}{K}\right)\left(1-\frac{A}{x}\right),\\
\frac{\pa g_2(x,y)}{\pa x}&=& \frac{\pa f_2 (x,u)}{\pa x}\;,\\
 \frac{\pa g_2(x,y)}{\pa y}&=&-1+\frac{\pa f_2 (x,u)}{\pa y}=-1\;.
 \end{eqnarray*}

\noindent We are now concerned with the case when $x=A$ and $y=f_2(A,y,u)$. 
For simplifications purposes, we put $a_0=f_1(A,u)$ and $b_0=f_2(A,u), \ds c_0:=\frac{\pa f_2(A,y,u)}{\pa x}$.
From above, the Jacobian matrix is given as 
\[ 
J(A,b_0)=\begin{pmatrix} 
2\frac{b_0}{A}\left(a_0-b_0\frac{A}{K}\right)  \vspace{0.2cm} & \ds 0\\
\ds- c_0& \ds -1
 \end{pmatrix}\;.\]
Since the matrix $J(A,b_0)$ is a lower triangular matrix, the eigenvalues are there $\ds \lambda_1=2\frac{b_0}{A}\left(a_0-b_0\frac{A}{K}\right)  $ and $\lambda_1=-1$.
Thus 
if  $b_0\left(b_0A-a_0K\right)<0$, then $\lambda_1<0$ and $\lambda_2<0$, so the point $(A, b_0)$ is an attractor.
 If  on the other hand  $b_0\left(a_0K-b_0A\right)\geq 0$, then $\lambda_1<0$ and $\lambda_2>0$, so the point $(A, b_0)$ is a repeller. 
 Finally, if $\left(a_0K- b_0A\right)=0$, then $\lambda_1<-1$ and $\lambda_2=0$ and then $(A,b_0)$ is a saddle point.
 \end{proof}
 
\noindent  Finally, we discuss the case when the steady states  are $(x,y)$ with $y=f_2(x,u)$ and $\ds f_1(x,u)=\frac{yx}{K}$. In this case, \begin{eqnarray*}
tr(J) & = &  \frac{\pa g_1(x,y)}{\pa x}+\frac{\pa g_2(x,y)}{\pa y}\\
&=& 2y\left( \frac{\pa f_1}{\pa x}-\frac{y}{K}\right)\left(1-\frac{A}{x}\right)-1+\frac{\pa f_2}{\pa y}\\
&=& \frac{2Kf_1}{x}\left( \frac{\pa f_1}{\pa x}-\frac{f_1}{x}\right)\left(1-\frac{A}{x}\right)-1+\frac{\pa f_2}{\pa y}\;,
\end{eqnarray*}
and 
\begin{eqnarray*}
det(J)&=& \frac{\pa g_1(x,y)}{\pa x}\frac{\pa g_2(x,y)}{\pa y}-\frac{\pa g_2(x,y)}{\pa x} \frac{\pa g_1(x,y)}{\pa y}\\
&=&  \frac{2Kf_1}{x}\left( \frac{\pa f_1}{\pa x}-\frac{f_1}{x}\right)\left(1-\frac{A}{x}\right)\left(-1+\frac{\pa f_2}{\pa y}\right)+2f_1 \frac{\pa f_2}{\pa x} \left(1-\frac{A}{x}\right)\;.
\end{eqnarray*}
The eigenvalues are 
\begin{eqnarray*}
\lambda_1&=& \frac{1}{2}\left[tr(J)-\sqrt{\Delta}\right]\\
\lambda_2&=& \frac{1}{2}\left[tr(J)+\sqrt{\Delta}\right]\;,
\end{eqnarray*}
where \begin{eqnarray*}
\Delta&=&[tr(J)]^2-4det(J)\\
&=& \left[\frac{2Kf_1}{x}\left( \frac{\pa f_1}{\pa x}-\frac{f_1}{x}\right)\left(1-\frac{A}{x}\right)-1+\frac{\pa f_2}{\pa y}\right]^2\\
&-&4\left[\frac{2Kf_1}{x}\left( \frac{\pa f_1}{\pa x}-\frac{f_1}{x}\right)\left(1-\frac{A}{x}\right)\left(-1+\frac{\pa f_2}{\pa y}\right)+2f_1 \frac{\pa f_2}{\pa x} \left(1-\frac{A}{x}\right)\right]\\
&=&  \left[\frac{2Kf_1}{x}\left( \frac{\pa f_1}{\pa x}-\frac{f_1}{x}\right)\left(1-\frac{A}{x}\right)+1-\frac{\pa f_2}{\pa y}\right]^2-8f_1\frac{\pa f_2}{\pa x}\left(1-\frac{A}{x}\right)\;.
\end{eqnarray*}
If $\Delta>0$ with $det(J)>0$ and $tr(J)>0$, then $\lambda_2>0$. In this case,  $\sqrt{\Delta}<\abs{tr(J)}=tr(J)$ and it follows that $\lambda_1>0$. Consequently,  we have a repeller (unstable point) at $(x,y)$ such that $y=f_2(x,y,u)$ and $\ds f_1(x,u)=\frac{yx}{K}$.\\
If $\Delta>0$ with $det(J)>0$ and $tr(J)<0$, then $\lambda_1<0$. In this case,  $\sqrt{\Delta}<\abs{tr(J)}=-tr(J)$ and thus $\lambda_2<0$. Consequently,   we have an attractor (stable point) at $(x,y)$ such that $y=f_2(x,y,u)$ and $\ds f_1(x,u)=\frac{yx}{K}$.\\
If $det(J)<0$, then $\Delta>0$. If,  in addition  $tr(J)>0$, then  $\lambda_2>0$. In this case $\sqrt{\Delta}>\abs{tr(J)}=tr(J)$ and it follows that $\lambda_1<0$. Consequently,  we have a saddle at $(x,y)$ such that $y=f_2(x,y,u)$ and $\ds f_1(x,u)=\frac{yx}{K}$.\\
If $det(J)<0$, then $\Delta>0$. If, in addition  $tr(J)<0$, then  $\lambda_1<0$. Since $\sqrt{\Delta}>\abs{tr(J)}=-tr(J)$. Therefore $\lambda_2>0$.  Consequently,  we have a saddle at  $(x,y)$ such that $y=f_2(x,y,u)$ and $\ds f_1(x,u)=\frac{yx}{K}$.\\
We observe in particular that if $x=A$, then $det(J)=0$ and one of the eigenvalues $\lambda_1$ or $\lambda_2$ is 0.
 The other eigenvalue  is positive if $\ds \frac{\pa f_2}{\pa y}>1$ and negative if $\ds \frac{\pa f_2}{\pa y}<1$. 
 \subsection{Important particular case}
 Let us discuss the linear  case when  $T_1({\bf  W,u})={\bf u}$ and $T_2({\bf W,u},v)={\bf W}^T\cdot {\bf u}=\norm{{\bf W}}\cdot \norm{\bf u}\cos(\theta)$, where $\theta$ is the angle between the vectors ${\bf W}$ and ${\bf u}$. Therefore ${\bf W}\cdot T_1({\bf  W,u})={\bf W}^T\cdot {\bf u}$. This leads to  a modified Hebbian rule which becomes the Oja rule if $A=0$. Using $x:=\norm{W}^2$, and $u=\norm{\bf u}\cos(\theta)$,  we will have $f_1(x,u)=f_2(x,u)=u\sqrt{x}$. Since $x$ and $y$ have  the same dimension, without loss of generality, we can let $\tau_x=\tau_y=1$.\\
 The steady states for the system \eqref{eqn:Alleerule3} are $(x,0)$ for $x>0$, $(A,u\sqrt{A})$ and $(K, u\sqrt{K})$.\\
 From theorem \ref{thm1} above, $(x,0)$, for $x>0$,  is always a spiral sink, which occurs only when $u=0$. \\
 As for the steady state $(A,u\sqrt{A})$, we have $a_0=f_1(A,u)=u\sqrt{A}, b_0=f_2(A,y,u)=u\sqrt{A}, c_0=\frac{f_2(A,y,u)}{\pa x}=\frac{u}{2\sqrt{A}}$, and $d_0=\frac{f_1(A,y,u)}{\pa y}=0$.\\
 \noindent It follows that $1+d_0>0$ and  $b_0\left(a_0K-b_0A\right)=b_0^2\left(K-A\right)$. Therefore, by Theorem \ref{thm2} above, $(A, u\sqrt{A})$ is a an attractor if $K<A$ and a saddle  if $K>A$.\\
 \noindent For the steady state $(K,u\sqrt{K})$, we have 
 \begin{eqnarray*}
 \ds tr(J)&=& \frac{2Kf_1}{x}\left( \frac{\pa f_1}{\pa x}-\frac{f_1}{x}\right)\left(1-\frac{A}{x}\right)-1+\frac{\pa f_2}{\pa y}\\
&=&  \frac{2Ku\sqrt{x}}{x}\left( \frac{u}{2\sqrt{x}}-\frac{u\sqrt{x}}{x}\right)\left(1-\frac{A}{x}\right)-1\\
&=& \left(\frac{Ku^2}{x}-\frac{2Ku^2}{x}\right)\left(1-\frac{A}{x}\right)-1\\
&=&-\left[1+\frac{Ku^2}{x}\left(1-\frac{A}{x}\right)\right]\;.
 \end{eqnarray*}
 And with $x=K$, we will have: 
 \[tr(J)=-\left[1+u^2\left(1-\frac{A}{K}\right)\right]\;.\]
 We also have: 
 \begin{eqnarray*}
 det(J)&=& \frac{2Kf_1}{x}\left( \frac{\pa f_1}{\pa x}-\frac{f_1}{x}\right)\left(1-\frac{A}{x}\right)\left(-1+\frac{\pa f_2}{\pa y}\right)+2f_1 \frac{\pa f_2}{\pa x} \left(1-\frac{A}{x}\right)\\
 &=& \frac{2Ku\sqrt{x}}{x}\left( \frac{u}{2\sqrt{x}}-\frac{u\sqrt{x}}{x}\right)\left(1-\frac{A}{x}\right)(-1)+2u\sqrt{x}\frac{u}{2\sqrt{x}}\left(1-\frac{A}{x}\right)\\
 &=& (-1)\frac{-Ku^2}{x}\left(1-\frac{A}{x}\right)+u^2\left(1-\frac{A}{x}\right)\\
 &=&u^2\left(1-\frac{A}{x}\right)\left(1+\frac{K}{x}\right)\;.
 \end{eqnarray*}
 And with $x=K$, we will have 
 \[det(J)=2u^2\left(1-\frac{A}{K}\right)\;.\]
 Also, 
 \begin{eqnarray*} 
 \Delta&=&  \left[\frac{2Kf_1}{x}\left( \frac{\pa f_1}{\pa x}-\frac{f_1}{x}\right)\left(1-\frac{A}{x}\right)+1-\frac{\pa f_2}{\pa y}\right]^2-8f_1\frac{\pa f_2}{\pa x}\left(1-\frac{A}{x}\right)\\
 &=& \left[\frac{-Ku^2}{x}\left(1-\frac{A}{x}\right)+1 \right]^2-8u\sqrt{x}\frac{u}{2\sqrt{x}}\left(1-\frac{A}{x}\right)\\
 &=&  \left[\frac{-Ku^2}{x}\left(1-\frac{A}{x}\right)+1 \right]^2-4u^2\left(1-\frac{A}{x}\right)\\
 &=& \left[1-\frac{Ku^2}{x}\left(1-\frac{A}{x}\right) \right]^2-4u^2\left(1-\frac{A}{x}\right)\\
 \end{eqnarray*}
 For $x=K$, we have 
 \[\Delta=\left[1-u^2\left(1-\frac{A}{K}\right) \right]^2-4u^2\left(1-\frac{A}{K}\right)\;.\]
 Putting $\alpha=u^2\left(1-\frac{A}{K}\right)$, then $\Delta=\alpha^2-6\alpha+1$. The roots are
 \begin{eqnarray*}
 \alpha_1&=& \frac{6-\sqrt{32}}{2}, \quad \alpha_2= \frac{6+\sqrt{32}}{2}\;.
 \end{eqnarray*}
 It follows that $\Delta>0$ if $\alpha<\alpha_1$ or $\alpha>\alpha_2$ and $\Delta<0$ if $\alpha_1<\alpha<\alpha_2$.
 From Theorem \ref{thm2} above,  we conclude that $x=K$ is either a saddle or an attractor.
 
\subsection{Appendix B: Stability analysis of System \eqref{eqn:Alleerule445}}
\subsubsection{Appendix $B1$}

The steady states are obtained when we are on the $x-,y$-, and $z$-isoclines. As above, on the $x$-isocline, we have either $y=0$, or $\ds f_1(x,u)-\frac{yx}{K}=0$, or $A=\norm{\bf W}^2$. On the $y$-isocline, we will have $y=f_2(x,y,z,u)$, and on the $z$-isocline, we will have $z=1-f_3(x,y,u)$.\\
Now we let the Jacobian matrix at a point $(x,y,z)$ be
 \[  \begin{aligned}
J& :=J(x,y,z)=\begin{pmatrix}  \ds \alpha_1&  \ds \alpha_2 & \ds \alpha_3\\
\ds \beta_2 &  \ds \beta_2& \ds \beta_3\\
 \ds  \gamma_1 & \ds  \gamma_2& \ds \gamma_3\\
 \end{pmatrix}
  \end{aligned}
 \;.\]
 where \begin{eqnarray*}
 \alpha_1=\ds \frac{\pa g_1(x,y,z)}{\pa x} &=& 2y\left(\frac{\pa f_1(x,u)}{\pa x}-\frac{y}{K}\right)\left(1-\frac{A}{x}\right)+2y\left( f_1(x,u)-\frac{yx}{K}\right)\left(\frac{A}{x^2}\right)\\
 \ds \alpha_2=\frac{\pa g_1(x,y,z)}{\pa y} &=& \ds \left( 2f_1(x,u)-4\frac{yx}{K}\right)\left(1-\frac{A}{x}\right), \quad  \ds \alpha_3=\frac{\pa g_1(x,y,z)}{\pa z}=0\\
  \ds \beta_1=\frac{\pa g_2(x,y,z)}{\pa x} &=& \ds \frac{\pa f_2(x,y,z,u)}{\pa x}, \quad     \ds \beta_2= \frac{\pa g_2(x,y,z)}{\pa y} = \ds-1+ \frac{\pa f_2(x,y,z,u)}{\pa y} \\
    \ds \beta_3 =\frac{\pa g_2(x,y,z)}{\pa z} &=& \ds \frac{\pa f_2(x,y,z,u)}{\pa z}, \quad     \ds \gamma_1= \frac{\pa g_3(x,y,z)}{\pa x} =  \ds  \frac{\pa f_3(x,y,u)}{\pa x} \\
    \ds \gamma_2= \frac{\pa g_3(x,y,z)}{\pa y} &=&  \ds  \frac{\pa f_3(x,y,u)}{\pa y}, \quad          \ds \gamma_3= \frac{\pa g_3(x,y,z)}{\pa z} =  \ds  -1\;.  
 \end{eqnarray*}
 In the linear  case, we have $f_1(x,u)=u\sqrt{x}, f_2(x,y,z,u)=u\sqrt{x}+zy$. With Goodall's model, we will have  $f_3(x,y,u)=-uy\sqrt{x}$. It follows that 
 \begin{eqnarray*}
 \ds \alpha_1&=& 2y\left(\frac{u}{2\sqrt{x}}-\frac{y}{K}\right)\left(1-\frac{A}{x}\right)+2y\left( u\sqrt{x}-\frac{2yx}{K}\right)\left(\frac{A}{x^2}\right)\\
 \alpha_2&=& \ds \left( 2u\sqrt{x}-4\frac{yx}{K}\right)\left(1-\frac{A}{x}\right),  \quad \quad \ds \alpha_3=0\\
  \ds \beta_1&=& \ds \frac{u}{2\sqrt{x}},\quad  \quad   \beta_2= \ds z-1,\quad \quad    \ds \beta_3=y\\
  \ds \gamma_1&=&  \ds  \frac{uy}{2\sqrt{x}}=y\beta_1,  \quad  \quad  \ds \gamma_2=-u\sqrt{x},  \quad  \quad    \ds \gamma_3=  \ds  -1\;. 
 \end{eqnarray*}
Now let us discuss the stability of the steady states.\\
 
 \noindent{\bf Case 1: Stability of the line $(x,0,1),~ x>0$}.\\
 In this case, we have
 \[A_0:=J(x,0,1)=\begin{pmatrix}  0 &  \ds 2u\sqrt{x} \left(1-\frac{A}{x}\right) & 0\\
\ds \frac{u}{2\sqrt{x}} &  0 & 0\\
0& -u\sqrt{x}& -1\\
 \end{pmatrix}\;.\]
 The eigenvalues are \[\lambda_1=-1, ~\lambda_2=u\sqrt{1-\frac{A}{x}}, ~\lambda_3=-u\sqrt{1-\frac{A}{x}}\;.\]
Since $y:=v=0\implies u=0$, the eigenvalues are actually   \[\lambda_1=-1,~\lambda_2=0,~ \lambda_3=0\;.\]
 Consequently,  the line$(x, 0,1)$ is always stable.\\
 \noindent{\bf Case 3: Stability of the points $\bm{B_i}, 1\leq i\leq 2$}.\\
 
  \noindent {\bf For $\bm{B_1}$}, we know that $x=A, y=-1$ and $\ds z=1+u\sqrt{A}$.  In this case,
     \begin{eqnarray*}
     \alpha_1&=& -2\rb{\frac{u}{\sqrt{A}}+\frac{2}{K}}; \quad \alpha_2=0,\quad \alpha_3=0\\
     \beta_1&=&\frac{u}{2\sqrt{A}}, \quad \beta_2=u\sqrt{A}; \quad \beta_3=-1\\
     \gamma_1&=& -\frac{u}{2\sqrt{A}}; \quad \gamma_2=-u\sqrt{A}, \quad \gamma_3=-1\;. 
     \end{eqnarray*}
   Reparametrizing as $\ds \beta=\frac{u}{2\sqrt{A}}$ and $\gamma=u\sqrt{A}$, the Jacobian matrix is given as 
    \[A_{31}:=\begin{pmatrix}  \alpha_1 &  0& 0\\
\beta& \gamma& -1\\
\ds -\beta & -\gamma& -1\\
 \end{pmatrix}\;.\]
 
  \[\lambda_1=\alpha_1, \quad  \lambda_2= \frac{1}{2}\left[ \gamma-1+\sqrt{(\gamma-1)^2+8\gamma}\right], \quad  \lambda_3= \frac{1}{2}\left[ \gamma-1-\sqrt{(\gamma-1)^2+8\gamma}\right]\;.\]

  \noindent {\bf Case 311: $u>0$}.\\
    In this case, $\gamma>0$, and therefore,  \[\sqrt{(\gamma-1)^2+8\gamma}=\sqrt{\gamma^2+6\gamma+1}=\sqrt{(\gamma+1)^2+4\gamma}\geq \abs{\gamma+1}=\gamma+1.\]
   It follows that  $2\lambda_2=\gamma-1+\sqrt{(\gamma+1)^2+4\gamma}\geq 2\gamma>0$, and consequently, the point $B_1$ is unstable.\\
   
   \noindent  {\bf Case 312: $u<0$}.\\
    In this case, $\gamma<0$, and therefore, $\lambda_3<0$. We also have  \[0\leq \sqrt{(\gamma-1)^2+8\gamma}\leq \abs{\gamma-1}.\]
    It follows that
    \[\gamma-1\leq 2\lambda_2\leq \gamma-1+\abs{\gamma-1}\;.\]
Since $\gamma<0<1$, we conclude that \[\gamma-1\leq 2\lambda_2\leq 0.\]

\noindent We finally note that $\lambda_1=\alpha_1\leq 0$ if $u\geq -\frac{2\sqrt{A}}{K}$. We note that 
\[\max\set{-\frac{2\sqrt{A}}{K},-\frac{2\sqrt{A}}{2A}}=-2\sqrt{A}\min\set{ \frac{1}{2A}, \frac{1}{K}}=-2\sqrt{A}\max\set{2A,K}\;.\]

We conclude that 

\noindent \[\mbox{if $-\frac{2\sqrt{A}}{K}<2\sqrt{A}\max\set{2A,K}<u<0$, the point $B_1$ is stable and if not, it is unstable}.\]

 \noindent {\bf For $\bm{B_2}$}, we know that $x=A, y=1$ and $\ds z=1-u\sqrt{A}$. In this case, we have 
 
  \begin{eqnarray*}
     \alpha_1&=& 2\rb{\frac{u}{\sqrt{A}}-\frac{2}{K}}; \quad \alpha_2=0,\quad \alpha_3=0\\
     \beta_1&=&\frac{u}{2\sqrt{A}}, \quad \beta_2=-u\sqrt{A}; \quad \beta_3=-1\\
     \gamma_1&=& -\frac{u}{2\sqrt{A}}; \quad \gamma_2=-u\sqrt{A}, \quad \gamma_3=-1\;. 
     \end{eqnarray*}

  The Jacobian matrix is given as 
    \[A_{32}:=\begin{pmatrix}  \alpha_1 &  0& 0\\
\beta_1& -\gamma& -1\\
\ds \beta_1 &-\gamma& -1\\
 \end{pmatrix}\;.\]
 We find the eigenvalues to be 
\[\lambda_1=0 \quad  \lambda_2=\alpha_1, \quad  \lambda_3= -\gamma-1\;.\]

   \noindent  {\bf Case 321: $u>0$}.\\
   In this case, $\lambda_3<0$ since  $\gamma>0$. If $\ds u<\frac{2\sqrt{A}}{K}$, then $\lambda_2=\alpha_1=2\rb{\frac{u}{\sqrt{A}}-\frac{2}{K}}<0$. It follows that    $B_2$ is stable, if not, it is unstable.
   
   \noindent  {\bf Case 322: $u<0$}.\\
   Then  $\lambda_3=-\gamma-1<0$ if $\gamma>-1$, that is, if $u\sqrt{A}>-1$. Also, $\lambda_2=\alpha_1<0$ if $u<0$.
   In conclusion
    \noindent \[\mbox{if $-\frac{1}{\sqrt{A}}=-\frac{2\sqrt{A}}{2A }<-2\sqrt{A}\max\set{2A,K}<u<0$, the point $B_2$ is stable}.\]

   \subsubsection{Appendix $B2$}
    
   \noindent{\bf Case 2: Stability of the lines $L_i, 1\leq i\leq 2$}.\\
   
   For $L_1$, we know that $y=1$ and $\ds z=1-\frac{x}{K}$, and 
   \begin{eqnarray*}
   \alpha_1&=&2\rb{\frac{u}{2\sqrt{x}}-\frac{1}{K}}\rb{1-\frac{1}{A}}+2\rb{u\sqrt{x}-\frac{2x}{K}}\rb{\frac{A}{x^2}} \\
   \alpha_2&=& \ds\rb{ 2u\sqrt{x}-\frac{4x}{K}} \left(1-\frac{A}{x}\right),\quad \quad \alpha_3=0\\
   \ds \beta_1&=&\frac{u}{2\sqrt{x}}\quad \quad \beta_2= -\frac{x}{K} \quad \quad \beta_3=-1\\
   \ds \gamma_1&=&\frac{u}{2\sqrt{x}}, \quad \quad \gamma_2=-u\sqrt{x}, \quad \quad \gamma_3=-1
   \end{eqnarray*}
   
   The Jacobian matrix is given as 
    \[A_{21}:=\begin{pmatrix}  \alpha_1& \alpha_2   & 0\\
\beta_1 & \beta_2& -1\\
\beta_1& \gamma_2& -1\\
 \end{pmatrix}\;.\]
   The characteristic polynomial is $P(-\lambda)=-\lambda^3+a_2\lambda^2+a_1\lambda+a_0$.
   where \begin{eqnarray*}
   a_2&=&(\alpha_1+\beta_2-1)\\
   a_1&=&\alpha_1(1-\beta_2)+\beta_2-\gamma_2+\alpha_2\beta_1\\
   a_0&=& \alpha_1(\gamma_2-\beta_2)
   \end{eqnarray*}
From Viete's theorem \cite{Viete1646}, there are three cases to consider:

 \noindent {\bf Case 21: $a_0=0$}.\\
 
 In this case, $P(\lambda)=\lambda(-\lambda^2+a_2\lambda+a_1)$. The roots are \
 \[\lambda_1=0, \quad \lambda_2=\frac{1}{2}\left[a_2-\sqrt{a_2^2+4a_1}\right],\quad  \lambda_3=\frac{1}{2}\left[a_2+\sqrt{a_2^2+4a_1}\right]\;.\]
Suppose that $a_2^2+4a_1<0$.  If $a_2<0$, then $\lambda_2$ and $\lambda_3$ are complex conjugates eigenvalues with negative real parts, thus  $L_1$ is stable attractor. If $a_2>0$, then  $L_1$  is unstable. \\
 
\noindent Suppose that $a_2^2+4a_1>0$.  If $a_1>0$ and $a_2>0$, then $\lambda_2$ and $\lambda_3$ are real eigenvalues with $\lambda_3>0$, thus $L_1$ is unstable.  If $a_1<0$ and $a_2<0$, then  $\lambda_2<0$ and $\lambda_3\leq \frac{1}{2}[a_2+|a_2|]=0$. It follows that $L_1$ is stable. If $a_1>0$ and $a_2<0$ or $a_1<0$ and $a_2>0$, then one of the eigenvalues if positive, and thus $L_1$ is unstable.\\
 
  \noindent {\bf Case 22: $a_0<0$}.\\
 
 This means that we have 3 cases: 3 negative eigenvalues, 1 negative eigenvalue and 2 complex conjugate eigenvalues, or 2 positive and 1 negative eigenvalues. Clearly, in case the eigenvalues are all negative, $L_1$ is stable. In the third case, $L_1$ is unstable. In the second case, $L_1$ may or may not be stable. 
 
   \noindent {\bf Case 22: $a_0>0$}.\\
 This means that we also have 3 cases: 3 positive roots, 1 positive eigenvalues and 2 complex conjugate eigenvalues, 2 negative eigenvalues and one positive eigenvalue. In the first  and third cases,  $L_1$ is unstable. In the second case, $L_1$ may or may not be stable. \\
 
  \noindent   For $L_2$, we know that $y=-1$ and $\ds z=1-\frac{x}{K}$. The only difference between this case and the previous is that here, \begin{eqnarray*}
  \alpha_1&=&-2\rb{\frac{u}{2\sqrt{x}}+\frac{1}{K}}\rb{1-\frac{1}{A}}+2\rb{u\sqrt{x}+\frac{2x}{K}}\rb{\frac{A}{x^2}}\\
\alpha_2&=&\ds-\rb{ 2u\sqrt{x}+\frac{4x}{K}} \left(1-\frac{A}{x}\right) .
  \end{eqnarray*}

 \end{document}